\documentclass[aps,prb,superscriptaddress,nofootinbib]{revtex4}
\usepackage{graphicx}
\usepackage{hyperref}
\usepackage{bm}

\begin{document}

\def\etal{\emph{et al.},~}
\def\Tc{$T_{\rm{c}}$}
\def\ScCuborate{SrCu$_2$(BO$_3$)$_2$}
\def\T1T{$T_1T$}
\def\YBCO{YBa$_2$Cu$_3$O$_{6+y}$}
\def\LSCO{La$_{2-x-y}$Nd$_y$Sr$_x$Cu$_{2}$O$_{4+\delta}$}
\def\Bi2212{Bi$_{2}$Sr$_{2}$CaCu$_{2}$O$_{8+\delta}$}
\def\Hg1201{HgBa$_{2}$CuO$_{4+\delta}$}
\def\Néel{N\'eel}
\def\Hc{$H_{\rm{c}}$~}
\def\Hs{$H_{\rm{s}}$}
\def\PRL{Phys. Rev. Lett.~}
\def\PRB{Phys. Rev.~B~}

\title{Nuclear Magnetic Resonance in High Magnetic Field: Application to Condensed Matter Physics}
\author{Claude Berthier},
\affiliation{Laboratoire National des Champs Magn\'etiques Intenses-CNRS (UPR 3228), EMFL, UGA, UPS, INSA, 38042 Grenoble, France}
\author{Mladen Horvati\'c},
\affiliation{Laboratoire National des Champs Magn\'etiques Intenses-CNRS (UPR 3228), EMFL, UGA, UPS, INSA, 38042 Grenoble, France}
\author{Marc-Henri Julien},
\affiliation{Laboratoire National des Champs Magn\'etiques Intenses-CNRS (UPR 3228), EMFL, UGA, UPS, INSA, 38042 Grenoble, France}
\author{Hadrien Mayaffre},
\affiliation{Laboratoire National des Champs Magn\'etiques Intenses-CNRS (UPR 3228), EMFL, UGA, UPS, INSA, 38042 Grenoble, France}
\author{Steffen Kr\"{a}mer}
\affiliation{Laboratoire National des Champs Magn\'etiques Intenses-CNRS (UPR 3228), EMFL, UGA, UPS, INSA, 38042 Grenoble, France}

\begin{abstract}
In this  review, we describe the potentialities offered by the
nuclear magnetic resonance (NMR) technique to explore at a microscopic level
 new quantum states of condensed matter induced by  high magnetic
fields. We  focus on experiments realised in resistive (up to 34~T) or hybrid (up to 45~T)
magnets, which open a large access to these quantum phase transitions. After an introduction on NMR observable, we consider several topics: quantum spin systems (spin-Peierls transition, spin ladders, spin nematic phases, magnetisation plateaus and Bose-Einstein condensation of triplet excitations), the field-induced charge density wave (CDW) in high $T_c$~superconductors, and exotic superconductivity including the Fulde-Ferrel-Larkin-Ovchinnikov superconducting state and the field-induced superconductivity due to the Jaccarino-Peter mechanism.

\end{abstract}
\maketitle

\section{\label{intro}Introduction}
 Since its discovery just after the second World War,
the nuclear magnetic resonance (NMR) technique  has known a
tremendous development in chemistry, biology and imaging for
medical applications (MRI). This development was founded on three
pillars: the development of superconducting magnets providing
extremely stable and homogeneous ($10^{-10}$) magnetic fields
up to 23.4 T (1 GHz for proton resonance), the continuously
increasing power of computers, and the development of high
frequency and high power  electronics. Most of the
experiments performed in the world concern structural information
and are usually performed around room temperature  (in particular for
biology and MRI) in diamagnetic systems. The situation is quite
different for NMR applied to solid state physics, where
temperature, pressure and magnetic field are essential
thermodynamic variables. The homogeneity and  stability requirements are
much less stringent than mentioned above, and usually fall in the
range $10^{-5}-10^{-3}$, depending on the systems under study, so
that in many cases all-purpose high-field resistive magnets, available only  in
few dedicated facilities in the world, can be used up to field values up to
35~T, or even 45~T in a hybrid magnet. In this case,
 high magnetic fields are not used to increase the sensitivity or the
resolution of NMR spectroscopy, but  as a
physical variable able to induce phase transitions, even at zero
temperature (the so-called quantum phase transitions) and to
access to new (quantum) phases of condensed matter [\onlinecite{Cargese}].
 Electrons in the matter couple to the magnetic field $H$
through their spin and orbit. In this latter case, it is
convenient to define a typical magnetic length $l_B =
\sqrt{\hbar/eB} = \frac{25\textrm{nm}}{\sqrt{B}[\textrm{T}]}$, such as $2\pi
l_B^2B=\phi_0=h/e$, where $e$ is the absolute value of the
electron charge, and $\phi_0$ the elementary flux quantum, and to
compare it to some typical distance of the system under study. The
most well known examples are the critical field $H_{c2}$  in a
superconductor of type II and the Integer Quantum Hall Effect
(IQHE) and the Fractional one (FQHE) in 2D electron gas. In the first case, the
comparison between the coherence length $\xi$ of the Cooper pairs
and the (superconducting) quantum flux $\phi_{0s}=h/(2e)$ gives
the upper critical field $H_{c2} = \phi_0/2\pi(\xi(T))^2 $ [\onlinecite{De Gennes}]. In the
second case, the IQHE plateaus correspond to incompressible phases
in which the number of electrons per flux quanta is an integer
[\onlinecite{Kitzing}]. A similar picture can be used for the FQHE
[\onlinecite{Stormer}] with composite fermions [\onlinecite{Jain,GlattliShankar}].
As far as the coupling with the spins are concerned, it is the
Zeeman energy which has to be compared with relevant energy scale
in the system under consideration. Examples range from quantum spin
systems, in which the characteristic energies derive from the exchange
couplings $J$'s, to the Pauli limit in superconductors when the
Zeeman energy overcomes the pairing energy of Cooper pairs. More
generally, application of magnetic field allows to generate new
quantum phases and, until recently, NMR has been the only
technique allowing a microscopic investigation of their structure
and excitations for field values above 17 T. This is now changing, with the new  possibilities for X-rays to do
experiments under pulsed magnetic fields up to 30~T
[\onlinecite{Xraypulsedfields}] and for neutron scattering up to 27~T
[\onlinecite{Berlin}]. Comparing results obtained by these techniques
with those obtained by NMR  opens a new fascinating area of research.

In this paper, we will review some of the NMR contributions to the
 physics in high magnetic fields performed by the authors [\onlinecite{FagotPRL,FagotPRB,ChaboussantPRL,Chaboussant_1998,HorvaticPRLCuGeO_99,Julien2000,Mladen_Cargese,Kodama2002,Takigawa2004,Kraemer2007,%
Hiraki_2007,Klanjsek_2008,Levy2008,Koutroulakis10,Tao2011,Mukhopadhyay_2012,Wu13,Tao2013,Takigawa2013,Kraemer2013,Jeong,Mayaffre14,Klanjsek_2015,%
Tao2015,Julien15,Iye,Jeong2016,Blinder2017,Orlova_2017,Zhou17,Zhou17b}]  using resistive
 magnets at the "Laboratoire National des Champs Magn\'etiques Intenses" (LNCMI-Grenoble). Some experiments requiring magnetic fields up to 45~T were performed at the National High Magnetic Field Laboratory (NHMFL) at Tallahasse
 (Florida, USA) and we  also  discuss recent NMR results obtained in pulsed magnetic field up to 55 T at the LNCMI-Toulouse.

\section{\label{NMRobs}NMR observables}
\label{NMR} Without entering into details of how  NMR is
actually performed [\onlinecite{Slichter,Abragam,Mehring,Nuts,NarathNMR}],
we will limit the presentation to its basic principles in order to
explain what physical quantities can be observed [\onlinecite{Mladen_Cargese}]. In a typical
configuration, NMR relies on the Zeeman interaction
\begin{equation}
\mathcal{H}_{\mathrm{Z}}= - \bm{\mu}_{\mathrm{n}}%
\cdot\mathbf{H}_{\mathrm{n}}
\end{equation} of the magnetic moment of
nuclei (of selected atomic species) $\mathbf{\mu}_{\mathrm{n}}=\hbar \gamma _{%
\mathrm{n}}\mathbf{I}_{\mathrm{n}}$, where $\gamma _{\mathrm{n}}$\ and $\mathbf{I}%
_{\mathrm{n}}$ are the gyromagnetic ratio and the spin of the nucleus, to obtain an information on the local magnetic field
value $\mathbf{H}_{\mathrm{n}} $ at this position. The experiment is performed in a magnetic field $H_{0}\sim 10$\thinspace T whose
value is precisely known (calibrated by NMR), and which is perfectly constant in time and homogeneous over the sample volume.
A resonance signal is observed at the Larmor frequency corresponding to transitions
between adjacent Zeeman energy levels $\omega _{\mathrm{NMR}}=\gamma _{%
\mathrm{n}}H_{\mathrm{n}}$, allowing very precise determination of $\mathbf{H}_{%
\mathrm{n}}$, and therefore of the local, induced, so-called
``hyperfine
field'' $\mathbf{H}_{\mathrm{hf}}=\mathbf{H}_{\mathrm{n}}-\mathbf{H}_{0}$\ (as $%
\gamma _{\mathrm{n}}$ is known from calibration on a convenient
reference sample). This hyperfine field, produced by the electrons
surrounding the chosen nuclear site, is a signature of
local electronic environment. On the other hand, nuclei which have
 a spin I $>$ 1/2  have a non-spherical distribution of
charge, and possess a quadrupolar moment which couples to the
electric field gradient (EFG) tensor produced by the surrounding electronic
and ionic charges. In a single crystal, the single NMR line corresponding to the Zeeman
interaction is then split into 2I lines, and this allows an accurate
determination of the EFG tensor, a quantity very sensitive to structural
transitions, or to a modulation of the electronic density, as
observed in CDW systems [\onlinecite{Berthier78}]. While the
NMR spectra correspond to static values (at the NMR scale) of the
hyperfine field and the EFG, the fluctuations of these quantities
are at the origin of the spin-lattice relaxation rate ($1/T_1$),
which measures their spectral density at the Larmor frequency.

\subsection{High Magnetic Field and NMR}

The spectral resolution of NMR is directly limited by the
temporal and spatial homogeneity of the external magnetic field
$H_{0}$. In the experiments where NMR is used for the
determination of complex molecular structures [\onlinecite {HighRes}], the
$H_{0}$\ field variations over the nominal sample dimension of
1\thinspace cm should be $\sim $\thinspace 10$^{-9}$\ for studies
in liquid solutions or 10$^{-6}$--10$^{-8}$\ in solid state
compounds. In both cases the magnetic field is produced by
commercially available, ``high-resolution'' superconducting (SC)
magnets, providing \textit{fixed} field, limited by the present SC
technology to a maximum field of 23.4\thinspace T. An SC magnet
operating at 28~T, using High $T_{\rm{c}}$ superconductor
technology, should be commercialized soon. The interest in high
fields for structural investigations is driven by the improvement of the resolution and the sensitivity.

When NMR is used as a probe of the electronic and magnetic
properties in solid state physics,
the required field homogeneity is typically much lower, 10$^{-3}$--10$^{-5}$%
, but the field should preferably be variable (sweepable). Up to
20\thinspace T such a field is available from commercial SC
``solid state NMR magnets'', with homogeneity of \linebreak
10$^{-5}$--10$^{-6}$. Higher fields (up to a maximum of
45\thinspace T) are available from big resistive or hybrid
(SC+resistive) magnets, but their homogeneity is not optimised for
NMR. Still, a typical value of 40$\times $10$^{-6}$ over a
1--2\thinspace mm sample positioned precisely in the field center
is satisfactory for a great majority of solid state NMR studies.
However, because of small sample size requirement and very high
running cost, one uses these big magnets only for
NMR studies of magnetic field dependent phenomena like field
induced phase transitions.

\subsection{Local static observables}

The general Hamilonian for a species of spin $\mathbf{I}$, gyromagnetic
ratio $\gamma_i$ and quadrupole moment Q in a solid placed in a
an external magnetic field $H_{\mathrm{0}}$ can be written as
\begin{equation}
\mathcal{H} = \mathcal{H}_{Zemann} + \mathcal{H}_{hyperfine} +
\mathcal{H}_{quadrupolar} + \mathcal{H}_{spin-spin},
\end{equation}
in which $\mathcal{H}_{Zemann}= \sum_i -\gamma_{i}\hbar
H_{\mathrm{0}}I_z$,~$\mathcal{H}_{spin-spin}$ corresponds to the
nuclear-nuclear spin interaction and $\mathcal{H}_{hyperfine} +
\mathcal{H}_{quadrupolar}$ will be defined below. For simplicity,
we shall only consider the most common case where
$\mathcal{H}_{Zemann}$ $\gg$ $
 \mathcal{H}_{hyperfine}, \mathcal{H}_{quadrupolar}$, and
$\mathcal{H}_{spin-spin}$. In that case, one only retains the
secular parts of the perturbative Hamiltonians, which commute with
$\mathcal{H}_{Zemann}$.

In absence of unpaired electrons in the system,
$\mathcal{H}_{hyperfine}$ resumes to the so-called chemical shift
[\onlinecite{Abragam}], usually neglected in most of the metallic and
magnetic systems, except in some of them like the organic
conductors, as discussed in section \ref{exoticSC}. In all other
cases, the hyperfine Hamiltonian is dominated by the coupling with
unpaired electrons, which for one electron $\mathbf{s}$ at a
distance $\mathbf{r}$ writes as
\begin{equation}
\mathcal{H}_{hyperfine.} = 2\mu_{\rm{B}}\gamma_n \hbar
\mathbf{I}\cdot[\frac{\mathbf{l}}{r^3} - \frac{\mathbf{s}}{r^3} +
3\frac{\mathbf{r}(\mathbf{s}\cdot\mathbf{r})}{r^5} +
\frac{8\pi}{3}\mathbf{s}\delta(\mathbf{r})].
\end{equation}
The orbital coupling $\mathbf{l}\cdot\mathbf{I}/r^3$ is usually neglected at the first order, since the orbital moment $\mathbf{l}$ is quenched by the crystal field, except
when the spin-orbit coupling cannot be neglected. However, it
contributes to the second order producing a paramagnetic orbital
shift, which has the same origin as the Van-Vleck susceptibility
whatever one deals with magnetic insulators [\onlinecite{AbgBln}] or
metallic systems [\onlinecite{Winter}]. The other terms are responsible
for the following contributions to the hyperfine shift: the anisotropic
dipolar one due to electrons with $l \neq 0$, the
isotropic contact one (due to "$s$" electrons),  and the
core-polarisation (isotropic and most of the times negative) due to
the polarisation of the inner closed  $s$-shell by the open $p$ or
$d$ shells [\onlinecite{AbgBln,Winter}].

In the absence of quadrupole coupling, the frequency of a line in an NMR
spectrum gives a direct access to the local magnetic field at the
position of the chosen nucleus. More precisely, we get an average
of the local field on the
time scale of the measuring process, $\sim $\thinspace 10--100\thinspace $%
\mu$s for solid state NMR. For systems with localised electronic
spins in particular, it is easy to see that the induced field is
linearly dependent on the spin polarisation of the nearest
electronic spin(s) [\onlinecite{NarathNMR}]
\begin{equation}
\omega _{\mathrm{NMR}}/\gamma _{\mathrm{n}}=\left|
\mathbf{H}_{0}+\left\langle
\mathbf{H}_{\mathrm{hf}}\right\rangle \right| =\left| \mathbf{H}_{0}+{\sum }%
_{k}\,-A_{\mathrm{n},k}\left\langle \mathbf{S}_{k}\right\rangle +\mathbf{C}_{%
\mathrm{n}}\right| \;.  \label{HFfield}
\end{equation}
This linear dependence defines the hyperfine coupling constant (tensor) $A_{%
\mathrm{n},k}$ of nucleus ``n'' to electronic spin $\left\langle \mathbf{S}%
_{k}\right\rangle $ at position $k$, while
$\mathbf{C}_{\mathrm{n}}$ accounts for quadratic (second order
orbital or van Vleck) contributions which are not sensitive to the
spin direction, as well as (generally much smaller) contribution
of other (unpolarised) closed-shell electrons. Hyperfine coupling
\index{hyperfine coupling} will be very different according to the
distance between nuclear and electronic spins:

\begin{itemize}
\item  On-site ($\mathrm{n}=k$) hyperfine coupling is strong, $A\sim 1$%
--100\thinspace T, and is approximately known for a given spin
configuration (standard reference is [\onlinecite{AbgBln}]).

\item  When the coupling is "transferred'' or `"supertransferred''
by an exchange process (i.e., due to overlap of wave functions)
from the first or second neighbour site, its value is generally
impossible to predict.

\item  For any distant spins ($\mathrm{n}\neq k$), there is also a
direct magnetic dipole coupling, which is precisely known for given geometry ($%
\propto \left| \mathbf{r}_{\mathrm{n}}-\mathbf{r}_{k}\right|
^{-3}$), and is generally smaller than the (super)transferred
hyperfine coupling.
\end{itemize}

The interaction Hamiltonian corresponding to hyperfine coupling is $\mathcal{%
H}_{\mathrm{n},k}=\hbar \gamma _{\mathrm{n}}\mathbf{I}_{\mathrm{n}}\cdot A_{%
\mathrm{n},k}\cdot \mathbf{S}_{k}$\thinspace , and the
experimentally measured ``magnetic hyperfine shift''
\index{magnetic hyperfine shift} $K$ is defined as the frequency
shift with respect to the reference:
\begin{equation}
K(T)\equiv \omega _{\mathrm{NMR}}/\left( \gamma _{\mathrm{n}}H_{0}\right) -1=%
{\sum }_{k}\,A_{\mathrm{n},k\,}\left( g_{k}\mu
_{\mathrm{B}}\right) ^{-1}\chi _{k}(T)+K_{0},  \label{shift}
\end{equation}
where $g_{k}$\ and $\chi _{k}$\ are the $g$-tensor and the
magnetic
susceptibility (per site\thinspace !) tensor of the $k$ spins, $\mu _{%
\mathrm{ B}}$\ the Bohr magneton and $K_{0}$\ the shift corresponding to $%
\mathbf{C}_{\mathrm{n}}$\ term in (\ref{HFfield}). $K$ is a tensor
whose
different components are obtained for different orientations of $\mathbf{H}_{0}$%
. Regarding the left-hand side of (\ref{shift}), we remark that
NMR spectra can be \textit{equivalently} obtained either in a
fixed external field $H_{0}
$\ as a function of frequency, or at a fixed frequency $\omega _{\mathrm{NMR}%
}$\ as a function of magnetic field. This latter configuration is
more convenient for very wide spectra, except when the physical
properties of the sample strongly vary with $H_{0}$. Equation
(\ref{shift}), which is equivalent to
$\mathcal{H}_{\mathrm{n},k}$\ or\ (\ref{HFfield}), indicates how
the $A$\ tensor can be measured by NMR: when the temperature
dependence of ``bulk'' magnetic susceptibility $\chi
_{\mathrm{macro}}$\ is dominated by the spatially homogeneous
contribution of a single spin species, $A$ is
calculated from the slope $\Delta K(T)/\Delta \chi _{\mathrm{macro}}(T)$%
\thinspace . If $\left\langle \mathbf{S}\right\rangle $\ is taken
to be a number, then $A$ is given in units of magnetic field; for
historical
reasons, the number that is usually declared is the ``hyperfine field'' = $%
A/g\mu_{\rm{B}}$ (in Gauss/$\mu _{\mathrm{B}}$).

One  application of the determination of the hyperfine field is
to obtain the true temperature and (or) the field dependence of the
magnetisation of the spin system, which can not always been
obtained from bulk measurements since they may be
dominated by the contribution of impurities at low temperature [\onlinecite{Mendels2000}]. Another very
important point is to determine whether the magnetization is
uniform, or distributed over a commensurate or incommensurate
structure, as we shall see later.

In the case of metals, the hyperfine interaction is responsible for
the Knight shift, which can be written in the general case of a
transition metal as [\onlinecite{Winter}]:
\begin{equation}
K = K_s + K_d + K_{orb} + K_{dipolar} +  K_{chem},
\end{equation}
where $K_s (K_d$) are  proportional to the density of state at
the Fermi level of $s$ or $s-p$ band ($d$-band) respectively,
$K_{orb}$ depends on the filling of the $d$-band and is
proportional to the Van-Vleck susceptibility, and the chemical
shift $K_{chem}$ is usually negligible. $K_{dipolar}$  exists only in presence of $p$ or $d$ bands, and depends on the symmetry of the lattice [\onlinecite{Winter}]. In 3-dimensional (3D) systems and in absence of phase transition, $K_s$ and $K_{orb}$ are usually $T$-independent, while $K_d$ is $T$-dependent.
There are many other situations where the Knight shift strongly
varies with the temperature: quasi-1D organic conductors,
itinerant antiferromagnets or ferromagnets, heavy fermions, which
we do not discuss here. However it is worth  to say a word on the
case of superconductors. Below $T_{\rm{c}}$, a gap opens in the density of state, so that the
Knight shift $K$ will vanish at low temperature (except for the orbital and the chemical
contribution). The $T$-dependence of $K$($T$) depends of the symmetry
of the order parameter. In the case of an $s$-wave singlet
superconductor, $K$  decreases  exponentially at sufficiently low
temperature, the residual constant shift being due to the orbital and the
chemical contribution. For clean $d$-wave superconductors, due
to the presence of nodes in the gap, a linear behaviour is
expected at low temperature, after removal of the above mentioned
residual contributions. For triplet superconductors, the situation is more
complex and depends on the precise symmetry of the order parameter
[\onlinecite{Mineev}].

Let us now introduce the quadrupolar interaction. Its Hamiltonian
is  the part of the electrostatic interaction between the nuclei
and the electrons, which describes the interaction of the
electric field gradient (EFG) traceless tensor $V_{\alpha\alpha}$
with the quadrupole moment $Q$ of the nuclei . It can be written in the
frame of the principal axes $X,Y,Z$ of the EFG tensor as
\begin{equation}
\mathcal{H}_{quadrupolar} =
\frac{eV_{ZZ}Q}{4I(2I-1)}[3I_Z^2-I(I+1) + \frac{1}{2}\eta(I_+^2 +
I_-^2)],
\end{equation}
where the axes are chosen to satisfy $|V_{ZZ}|\geq |V_{YY}|\geq |V_{XX}|$,
 which constrains the asymmetry parameter $\eta$=$\frac{V_{YY}-V_{XX}}{V_{ZZ}}$ to $0 \leq\eta\leq1$. If  the quadrupolar coupling can be treated as a
perturbation, it is more convenient to express this Hamiltonian in
the laboratory frame where the  quantization is along the direction $z$ of the
applied field. For simplicity, we assume that $\eta = 0$, and
limit ourselves to the first order in perturbation. We can rewrite
the Hamiltonian as
\begin{equation}
\mathcal{H} =
\frac{h\nu_Q}{6}[\frac{1}{2}(3\cos^2\theta-1)(3I_z^2-I(I+1)],
\end{equation}
where $\nu_Q = 3eQV_{ZZ}/h2I(2I-1)$, $\theta$ is the angle between
the $Z$ direction and the applied magnetic field $H_0$ and $I\geq
1$. In a single crystal, the single line whose position is defined by the
Zeeman coupling $\gamma\hbar H_0(1+K)I_z$ is split into $2I$
equidistant components, separated by $\nu_Q(3\cos^2\theta-1)/2$.  In powder samples,  these 2I lines are thus distributed over a frequency range spanning  $(2I-1)\nu_{\rm{Q}}$.  For half-integer spins, there is a central line (corresponding to the (-1/2, 1/2) transition), which is not  affected to the first order, while for integer spins  there is no central line left.
 Note that nuclei can have several isotopes of different natural abundance, with different gyromagnetic ratio and different quadrupole moment. For example, a single Cu site (spin 3/2) with a specific electronic spin polarisation and a specific charge environment, will give rise to six different lines, 3 for $^{63}$Cu and 3 for $^{65}$Cu (see section 3.3).

It is easy to see that a distribution of quadrupolar couplings, due to disorder or to the presence a CDW
[\onlinecite{Berthier78,Butaud}] will change the shape of the satellites lines by an order of magnitude larger than that of the
central line. The shape of the satellite lines will crucially depend on
the symmetry of the CDW, the number of the wave-vectors defining the modulation ($1q,2q,3q$),
and of its commensurate or incommensurate character.

One also notices that a spin-density wave  (SDW) induces a modulation of $K$ which usually is easily distinguished from a CDW since it will affect in the same way the central line and the satellites. However, a charge modulation also implies a
modulation of $K$ (which is important in the study of organic conductors undergoing a charge-ordering transition [\onlinecite{Hiraki}], where the nuclei under study are usually the $^{13}$C ( spin 1/2, $Q = 0$)).
 This effect on the spectrum is usually smaller than the associated quadrupolar perturbation, but it may happen that they are of the same order of magnitude, like in underdoped YBa$_{2}$Cu$_{3}$O$_{6+x}$ discussed in section 4. In that case, the two phenomena can still be disentangled, but in a more subtle way.

To conclude this section on static NMR observables, it is important to say that in the most general case, where the symmetry is lower
that tetragonal or trigonal, the general form of the Hamiltonian is more complicated than mentioned here. It depends on $\theta$ and
$\phi$ which define the orientation of the magnetic field with respect to the main axes of the quadrupolar tensor, and for strong quadrupolar couplings it may also be necessary to  fit the spectra to the results of a fully diagonalised Hamiltonian to determine accurately $\nu_q$, $\eta$
and $K(\theta,\phi)$.

\subsection{Dynamic observables}
There are essentially two dynamic quantities used in NMR applied to solid state physics, which are  the
spin-lattice relaxation rate $1/T_1$, and the spin-spin relaxation
 rate $1/T_2$. In the absence of static magnetic field or EFG gradients inhomogeneities, $T_2$ is merely the correlation time of the transverse magnetization. Most of the time and in all the experiments described here, the time decay of the transverse magnetization is dominated by the inhomogeneities, and spectra, $1/T_1$ and $1/T_2$ are measured using the spin-echo technique [\onlinecite{Slichter,Abragam,Mehring,Nuts,NarathNMR,Mladen_Cargese}]. Since we do not use $1/T_2$ in the following, we shall not say more on this quantity (see [\onlinecite{Mladen_Cargese}] for further information).  $1/T_1$ measures the characteristic time for the
longitudinal nuclear magnetisation $M_{z}$ ($z$$\parallel$$H_{0}$)
to return to its thermal equilibrium value after a perturbation,
which is usually a destruction or an inversion of $M_{z}$. It
essentially measures the weight of the spectral density at the
Larmor frequency of the time fluctuations of the local hyperfine
field or of the quadrupole couplings. In the following, we shall consider only the fluctuations of the hyperfine field.  In localised spin systems, the part of the hyperfine field of interest for the relaxation can be written as:
$\delta \mathbf{h}(t) = \mathbf{h}(t) - <\mathbf{h}(t)> $ where
$<\mathbf{h}(t)>  $ is the time-average value at the NMR time scale. For an applied field along the $z$ direction,
\begin{equation}
1/T_{1z} = \frac{1}{2}\gamma_n^2 \int_{-\infty}^{\infty}d\textrm{t}~
\rm{e}^{i\omega_{NMR}t}\langle \delta \mathbf{h}_{+}(t) \delta
\mathbf{h}_{-}(0)\rangle
\end{equation}
which can be explicitly expanded as :
\begin{equation}
T_{1z}^{-1} = \frac{1}{2}\gamma _{\mathrm{n}}^{2}\int_{-\infty
}^{\infty }d\textrm{t}~\rm{e}^{i\omega_{NMR}t}
[(A_{xz}^{2}+A_{yz}^{2})\,\langle S_{z}(t)S_{z}(0)\rangle
+(A_{xx}^{2}+A_{yx}^{2})\langle S_{x}(t)S_{x}(0)\rangle
+(A_{yy}^{2}+A_{xy}^{2})\langle S_{y}(t)S_{y}(0)\rangle]
\end{equation}
Note that for a hyperfine coupling tensor $A$ that is \emph{non}-diagonal (in the laboratory frame) both parallel and
transverse (to external field $H_{0}$$\parallel$$z$) spin--spin correlation functions contribute to the relaxation, while only the
latter contribution is active if $A$ is diagonal.
In general, $A_{\alpha \neq \beta }\neq 0$\ as soon as $H_{0}$\ is \textit{%
not} parallel to a principal axis of the $A$ tensor,  which leads to a
complicated angular dependence of $T_{1}$. This introduces the  longitudinal  correlation $\langle S_{z}(t)S_{z}(0)\rangle$ function in the calculation of 1/$T_1$, which usually involves different relaxation mechanisms than those associated to the transverse one [\onlinecite{Mladen_Cargese,ChaboussantPRL}].

To take into account the coupling to several electronic spins we replace $%
A\cdot \vec{S}$\ by $\sum_{\vec{r}}A(\vec{r})\cdot \vec{S}%
(\vec{r})=\sum_{\vec{q}}\,A(\vec{q})\cdot \vec{S}(-\vec{q})$ to
get (assuming inversion and translation symmetry):
\begin{equation}
T_{1z}^{-1}=\frac{1}{2}\gamma _{\mathrm{n}}^{2}\sum_{\textbf{q}}
\sum_{\alpha =x,y,z}( A_{x\alpha }^{2}(\textbf{q})+A_{y\alpha
}^{2}(\textbf{q}))\int_{-\infty}^{\infty}d\textrm{t}\rm{e}^{i\omega _{%
\mathrm{NMR}}t} \langle S_{\alpha }(\mathbf{q},t)S_{\alpha }(-\mathbf{q}%
,0)\rangle
\end{equation}

  A more general expression, which can also be used in the case of itinerant
electronic systems, is obtained using the  fluctuation-dissipation theorem:
\begin{equation}
T_{1z}^{-1}= k_{\mathrm{B}}T \gamma _{\mathrm{n}}^{2}
\sum_{\mathbf{q}}\sum_{\alpha = x,y,z}%
(A_{x \alpha }^{2}(\mathbf{q})+A_{y \alpha}%
^{2}(\mathbf{q}))\chi _{\alpha \alpha }^{\prime \prime }(%
\mathbf{q},\omega _{\mathrm{NMR}})/(\omega%
_{\mathrm{NMR}}g_{\alpha \alpha }^{2}\mu_{\mathrm{B}}^{2})
\end{equation}
In the case of simple metals with a single conduction band, the relaxation rate is often expressed as proportional to the square of the density of
states at the Fermi Level $N$($E_F$) [\onlinecite{Slichter,Winter}]. Since the Knight shift if also proportional to $N$($E_F$),
 this leads to the famous Korringa relationship:
\begin{equation}
K^2T_1T = \frac{\hbar}{4\pi k_{\rm{B}}} (\frac{\gamma_e}{\gamma_n})^2 = S_0,
\end{equation}
where $\gamma_e$ ($\gamma_n$) are respectively  the electron (nuclear) gyromagnetic ratios and  $k_{\rm{B}}$ the Boltzman constant.
The ratio $K^2T_1T$
 is used as an indicator of the dominant fluctuations in the system, since $K$ is proportional to $\chi\rm{(}\emph{q}=0\rm{)}$, while  $T_1T$ is proportional to $\omega_n/ {\sum_q}\chi"\rm{(}\emph{q},\omega_n\rm{)}$. In absence of electron-electron interaction, $\chi"\rm{(}\emph{q},\omega_n\rm{)}$ is flat and $\mathfrak{S} = K^2T_1T/S_0 =1$. In presence of ferromagnetic fluctuations, $\chi"\rm{(}\emph{q},\omega_n\rm{)}$ it is peaked at $q$ = 0, so that $\mathfrak{S}$ is smaller than 1, while in presence of antiferromagnetic (AF) fluctuations, it is larger than 1.
 The situation is more complicated in transition metals, since $s-p$ band and $d$-band have to be treated separately. Moreover,  within the $d$-band,
 the relaxation due to the core-polarization,  dipolar or anisotropic interactions, and the orbital relaxation have to be treated separately [\onlinecite{NarathNMR,Winter}].
  $1/T_1$ is also very important to characterise superconductors. In $s$-wave superconductors, a gap opens on the whole Fermi surface,
  so that the relaxation rate decreases exponentially at low temperature. Just below $T_{\textrm{c}}$, due to a divergence of the density of states and the so-called coherence factor of the BCS theory, one  should in principle observe the so-called Hebel-Slichter peak [\onlinecite{HebelSlichter}] in the relaxation rate. This peak is a hallmark of a singlet $s$-type  superconductor, but its non-observation doesn't mean that the order parameter is not $s$-like. This peak is absent in $p$ or $d$
  type superconductors [\onlinecite{Mineev}]. For $d$-type singlet pairing, the gap vanishes at nodes, around which the density of states is linear
  as a function of the energy. As a consequence, the Knight shift varies linearly as a function of $T$, while $\rm{(}$$T_1T$$\rm{)}^{-1}$ is proportional to $T^2$.
  In cuprate high \Tc~superconductors, which are $d$-wave, this behavior is usually difficult to observe, due to impurities. In quasi-2D organic conductors,
   which are very pure, this behaviour can be observed when the field is parallel to the superconducting planes. In that case, there are no more Abrikosov vortices, so the contribution of their cores to the relaxation disappears.  Only Josephson vortices are present, which do not contribute to $1/T_1$ [\onlinecite{Mayaffre95}] (see also section 5).

\section{\label{QSS}Quantum spin systems under applied magnetic field}
\subsection{Introduction}
 Although a spin is by essence a quantum object, the denomination  ``quantum spin systems" (QSS) is usually dedicated to systems of localized electronic spins having small spin values (1/2, 1, ... , for which the eigenvalues $S(S+1)$ of $\overrightarrow{S}^2$  strongly differ
 from $S^2$) and which are  dominantly coupled by AF exchange interactions. In low-dimensional systems,  thermal and quantum fluctuations are enhanced, and can destabilize the semi-classical long-range-order ground states of N\'eel type [\onlinecite{Auerbach,Schollwock,Giamarchi_book,Sachdev_book,Frustration_book_2011}]. Although the 1D spin chains have been studied very early by theorists, they are still an important playground to study experimentally modern concepts in magnetism, like, for example, quantum criticality, fractional excitations (spinons) and topological order. Moreover, since 1D spin systems can be mapped onto 1D strongly interacting spinless fermions through the Jordan-Wigner transformation [\onlinecite{Jordan_Wigner}], they allow, at the difference of other 1D systems like organic conductors, nanotubes or quantum wires, an accurate verification of the predictions of the Tomonaga-Luttinger liquid (TLL) \footnote{In a TLL,  all the correlation functions decay as power laws. A TLL of spinless fermions is characterised by two parameters $u$ and $K$, where $u$ is the velocity of the excitations, and $K$ a number which allows to calculate the exponents associated to the various correlation functions.} model  [\onlinecite{Giamarchi_book}], starting from the microscopic Hamiltonian [\onlinecite{Klanjsek_2008}].  Concerning quasi-2D spin systems,  high \Tc~superconductors  have promoted the search for exotic quantum ground states, notably after the suggestion by P.W. Anderson that they could be described as ``a resonant valence bond (RVB) systems" doped by holes or electrons [\onlinecite{Anderson}].
The purpose of this paper is to show how important is the magnetic field in
the physics of QSS, and to concentrate on their microscopic properties  and low-energy excitations.  Several techniques like neutron scattering [\onlinecite{Broholm}], EPR [\onlinecite{Motokawa}], and NMR [\onlinecite{Mladen_Cargese}] can give access to the microscopic properties of these new states, the most powerful being neutron scattering.  This technique is however presently limited to magnetic fields lower than 17 T, even though inelastic neutron scattering experiments up to 27 T should become available soon [\onlinecite{Berlin}]. This section is limited to NMR in QSS under high magnetic fields, and to experiments on systems in which the magnetic field plays a key role, as a parameter of the phase diagram or the spin-dynamics. This will cover NMR experiments
in quasi-1D spin chains and spin ladders  [\onlinecite{Schollwock,Giamarchi_book}], magnetization plateaus [\onlinecite{Takigawa_11}] and Bose-Einstein condensation (BEC) in coupled dimer systems [\onlinecite{Giamarchi_natphys,Zapf2014}].

     The general Hamiltonian of QSS  can be written as $\mathcal{H} = \sum_{i,j} \overrightarrow{S}_{i} \mathbf{J}_{ij}
\overrightarrow{S}_{j}  +\mathcal{H}_{pert}$, where $\mathbf{J}_{ij}$, the  exchange interaction (in most of
the cases AF) between spins $\overrightarrow{S}_{i}$, is taken to be symmetric and can thus be written as
 $\mathbf{J}_{ij}=\sum_{\alpha} J_{ij}^{\alpha \alpha}$. The perturbation Hamiltonian $\mathcal{H}_{pert}$ can correspond
to the Dzyaloshinskii-Moriya (DM) interaction
[\onlinecite{Dzyaloshinskii,Moriya}]  or staggered $g$-tensors, which both correspond to antisymmetric operators mixing the eingenstates of $S^2$, to the spin-phonon interaction, and (or) to the spin-spin dipolar interaction. In addition to the semi-classical ground states of
N\'eel type, two categories of ground states have to be distinguished. The basic element of both of them is the so-called
valence bond (VB) state, that is a pair of spin 1/2 forming a singlet state. The first family is formed by  systems which can be separated
in two distinct sublattices (bipartite systems). The ground state is  usually a Valence Bond Crystals (VBC), in which there is no long range order (LRO) for the spins, but there is one for the valence bonds. Typical cases are spin ladders with strong rung couplings, and more generally weakly
coupled spin dimer systems. The second one gathers the geometrically frustrated systems, the archetype of which being the 2D kagome lattice, which are expected in some case to have a spin liquid ground state [\onlinecite{Frustration_book_2011,Balents}]. In such a state, there is no longer any LRO, neither for the spins nor for the valence bonds, but the ground state is a quantum superposition of many valence bonds configurations. In this section we will mostly consider the systems in which the ground state in zero field is a collective singlet state, separated from the first excited triplet states by an energy gap $\Delta(0)$ (at $H = 0$) which can be closed at a magnetic field \Hc $= \Delta(0)/g\mu_{\textrm{B}}$.  We shall distinguish quasi-1D systems, like spin chains and spin ladders, which by using the Jordan-Wigner transformation [\onlinecite{Jordan_Wigner}] can be mapped on interacting spinless fermions and,  in most of the cases,  described in the framework of Tomonaga-Luttinger liquid (TLL), and quasi-2D or 3D weakly coupled dimers in
which the triplet excitations are rather described as hardcore bosons (or "triplons"), for which the applied magnetic field plays the role of the bare chemical potential [\onlinecite{Giamarchi_natphys}]. These triplons can undergo a BEC  [\onlinecite{Giamarchi_natphys,GiamTsvelik1999,Nikuni2000}] when their kinetic energy dominates their repulsion, while in the inverse situation, the triplons crystallise into magnetization plateaus for commensurate values of
the triplets density [\onlinecite{Takigawa_11,Rice2002}].

\subsection{Quasi-1D systems}
In the absence of perturbation, the spin 1/2 chains can be described by the XXZ Hamiltonian

$\mathcal{H} = \sum_{i} J\{\epsilon (S_{i}^{x}S_{i+1}^{x}+%
S_{i}^{y}S_{i+1}^{y})+ S_{i}^{z}S_{i+1}^{z}\}$. For $\epsilon \geq 1$, there is no gap in the low-energy excitations at all values of the magnetic field and the spin-spin correlation functions $\langle
S_{+}(0,0)S_{-}(R,t)\rangle$ (transverse) and $\langle
S_{z}(0,0)S_{z}(R,t)\rangle$ (longitudinal) decay as power laws with exponents $\eta_x$ and $\eta_z$. The transverse correlation function, which decays more slowly than the longitudinal one $\eta_x <  \eta_z$, becomes dominant for low energy properties.  If $\epsilon < 1$, the ground state is of \Néel type and separated from the excited states by a gap $\Delta(0)$, and can be closed at $H = H_{\textbf{c}}$. For $H \geq H_{\textbf{c}}$, the system enters into a TLL phase in which the longitudinal correlations are dominant at low energy [\onlinecite{Haldane,Kimura_2008b}]. However, increasing the magnetic field (i.e. the filling of the spinless fermions band), the exponents $\eta$ governing the decay of  both correlation functions cross, and above a second critical field $H^{*}_{c}$, the transverse one becomes dominant.

 Due to the inter-chain couplings  that are always present in real compounds, these systems enter into a 3D ordered state at finite temperature. For $\epsilon \geq 1$, between $H = 0$  and the saturation field $H_{s}$,  this is a canted transverse AF state, while for $\epsilon < 1$ this is a longitudinal, incommensurate, spin density wave between \Hc and $H^{*}$, followed by a canted AF state between $H^{*}$ and the saturation field $H_{s}$. As an example, the spin system BaCo$_2$V$_2$O$_8$  has been the subject of intense investigation these last years [\onlinecite{Kimura_2007,Kimura_2008,Canevet_2013,Grenier_2015,Klanjsek_2015}].

In the presence of a perturbation, like an alternation of the exchange coupling, a frustrating nearest neighbour coupling, or a spin-phonon coupling, the spin-chains become gapped. As described above, the application of a magnetic field can close the gap and allows the system to enter into a TLL regime. As an example requiring the use of resistive magnet for NMR investigation, we describe below the case of the spin-Peierls compound CuGeO$_3$. We shall also  mention the case of spin ladders, and that of the frustrated spin 1/2 chain LiCuVO$_4$ which is expected to present a nematic phase at high magnetic field around 45-50 T.

\subsubsection{The High Field Phase of the spin-Peierls compound
CuGeO$_{3}$}


\begin{figure}[h]%
\begin{minipage}{0.98\hsize}
\centerline{\includegraphics [scale = 0.8]{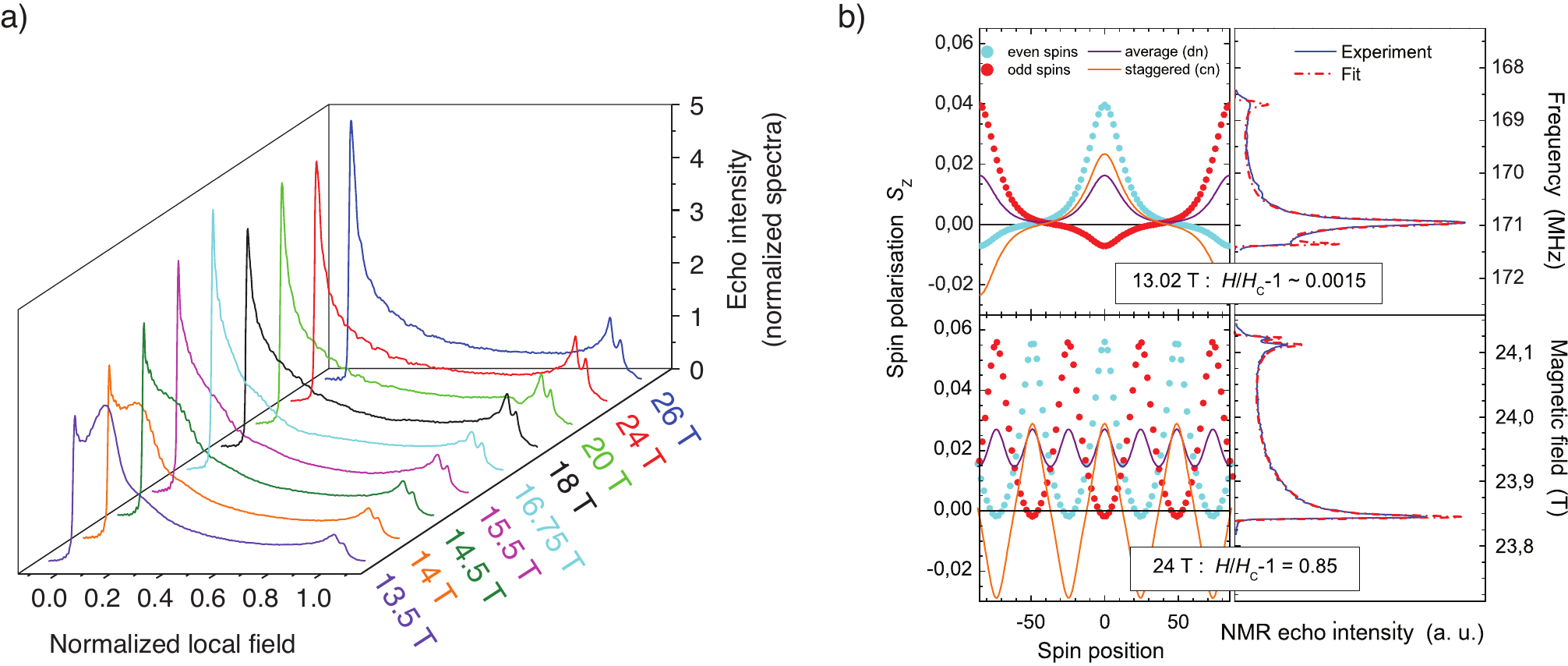}}
\begin{minipage}{0.98\hsize}
 \vspace*{-0.1cm}
 \caption[]{\label{Fig1}(Color online) (a)  Evolution of the Cu NMR lineshape
 as a function of the applied magnetic field in CuGeO$_{3}$ above the critical field $H_{\textrm{c}}$.
(b) \emph{left panels}: Reconstruction of the real space
spin-polarisation profile for $H$ = 13.02 and 24~T using Jacobi elliptic functions [\onlinecite{HorvaticPRLCuGeO_99}]. \emph{right panels}: The
corresponding fit (\emph{dashed line}) superposed on experimental NMR lineshape
(\emph{ full line}), taken in the
incommensurate high magnetic field phase of the spin--Peierls compound CuGeO$_{3}$. In this
way, NMR lineshape has provided the precise magnetic field dependence of the average and staggered spin-polarisation and the
magnetic correlation length up to 2\Hc [\onlinecite{HorvaticPRLCuGeO_99}].}
 \vspace*{0.1cm}
\end{minipage}
\end{minipage}
\end{figure}

A spin--Peierls chain is a Heisenberg, AF, $S=1/2$ chain on an elastic lattice, in which the exchange interaction coupling depends on the position of the magnetic atoms, which can vary to minimize the total energy [\onlinecite{Bray}]. At low temperature, this spin chain can gain energy by spontaneous dimerisation (deformation) of the lattice, which allows the opening of a gap in the low-energy magnetic excitations (absent in simple Heisenberg half-integer spin chains). This dimerised phase has a non-magnetic collective singlet ground state and an energy gap towards triplet excitations. Application of magnetic field
reduces the gap and, above the critical field $H_{\textrm{c}}$, drives the system into a magnetic phase with spatially inhomogeneous magnetisation (Fig.~\ref{Fig1}). In this field-induced phase, magnetisation appears as an incommensurate (IC) lattice of magnetisation peaks (solitons), where each soliton is bearing a total spin 1/2. The most studied spin--Peierls system is the inorganic compound CuGeO$_{3}$ [\onlinecite{Hase}], presenting a
spin--Peierls transition at 14--10\thinspace K (depending on $H$), and a critical field $\simeq 13$\emph{\thinspace }T. The NMR in
CuGeO$_{3}$ has been performed on the ``on-site'' copper nuclei which are directly and very strongly coupled to the electronic spin. In the high-temperature and in the dimerised phase, symmetric NMR lines are observed, reflecting spatially uniform magnetisation. Comparing $K(T)$ vs. $\chi _{\mathrm{macro}}(T)$
the complete hyperfine coupling tensor as well as the orbital
shift tensor $K_{0}=K(T=0)$ could be determined, and were found in good agreement with the values expected for a $d_{X^{2}-Y^{2}}$ orbital of Cu$^{++}$ ion [\onlinecite{AbgBln,FagotPRB}].

Above \Hc each line is converted to a very wide asymmetric spectrum (Fig.~\ref{Fig1}) corresponding to a spatially non-uniform distribution of magnetisation. This NMR lineshape is in fact the density distribution of the local magnetisation, i.e., of the spin polarisation. Therefore, for a periodic function in 1D, it can be directly converted to the corresponding real-space spin-polarisation profile (soliton
lattice, shown in Fig.~\ref{Fig1} \mbox{[\onlinecite{FagotPRL,HorvaticPRLCuGeO_99}]}. It was thus possible to obtain a full \textit{%
quantitative} description of the $H$ dependence of the spin-polarisation profile in the range from \Hc to 2\Hc= 26$\thinspace$T [\onlinecite{HorvaticPRLCuGeO_99}], clearly showing how the modulation of magnetisation
evolves from the limit of nearly independent solitons just above $H_{\mathrm{c}}$, up to the high magnetic field limit, where it becomes nearly
sinusoidal. Analysis of these data proved that the staggered component of magnetisation is reduced in the NMR image by phason-type motion of the soliton lattice [\onlinecite{Uhrig,Ronnow}]. The magnetic correlation length is found to be smaller than that associated to the lattice deformation (measured by X-rays
[\onlinecite{Kiryhukin}]), which is a direct consequence of the frustration due to the second neighbour interaction in the system [\onlinecite{Uhrig}]. The observed field dependence of the correlation length remains to be understood.

\subsubsection{Spin ladders}
\begin{figure}[t]
\vspace*{-1.1cm}
\begin{minipage}{0.98\hsize}
\centerline{\includegraphics[scale=0.45]{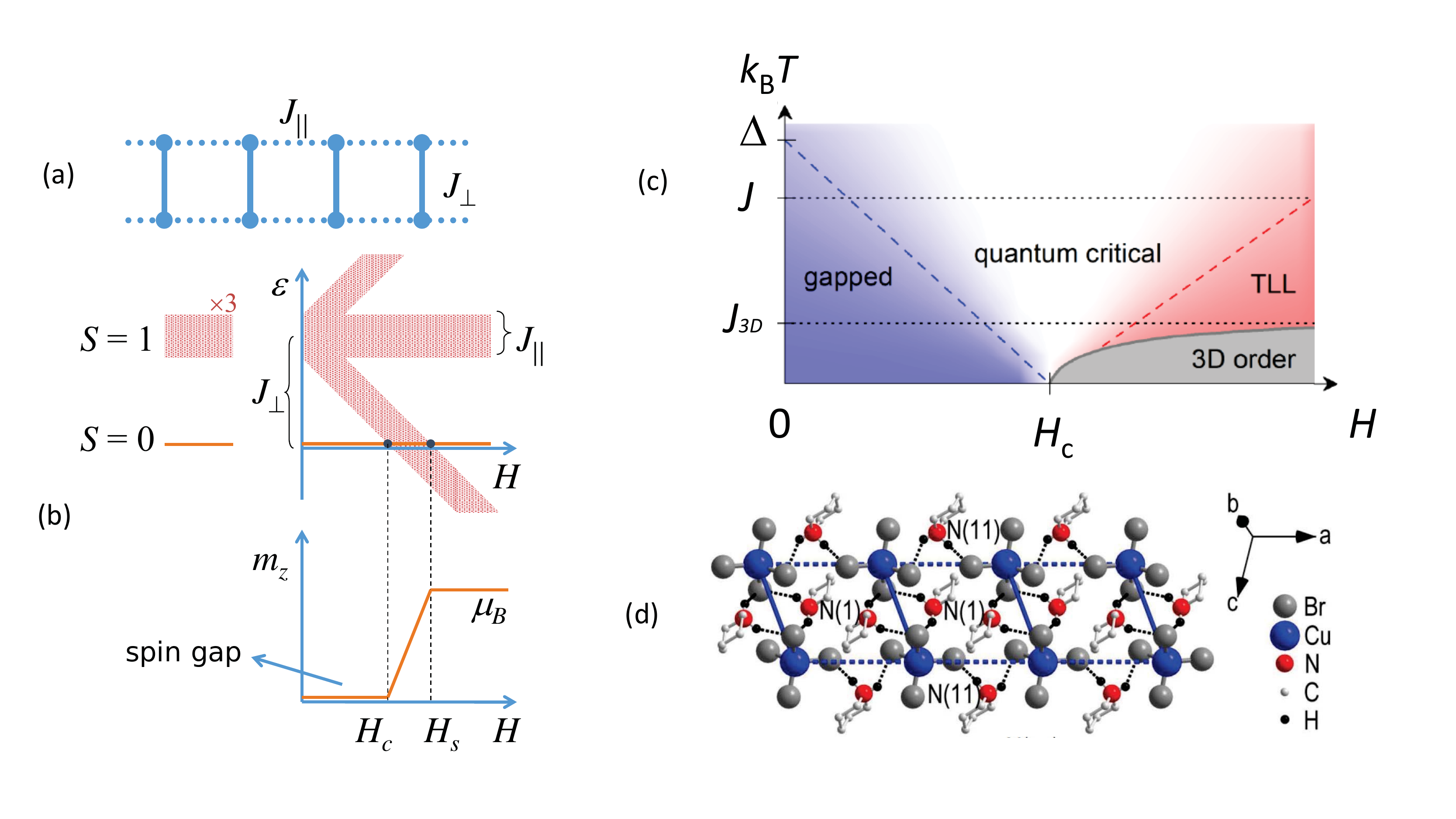}} 
\begin{minipage}{.98\hsize}
\caption[]{\label{Fig_spin_ladder_phase_diagram_v2} (Color online)   (a) Schematic representation of a simple spin ladder . (b) Schematic representation of the energy levels as a function of the applied magnetic field $H$ in the case of a strong rung coupling. For a single dimer (rung), the first triplet excitation is separated from the singlet ground state by a gap $\Delta(H=0) = J_{\perp}$. Due to the  interactions $J_{\parallel}$ along the legs, the triplet excitations form a band, which splits into three under the application of  $H$. The lowest band first crosses the collective singlet ground state at the field $H_{\rm{c}}$ (closure of the gap). This corresponds to the first quantum critical point [\onlinecite{Sachdev_book}], at which the ground state switches from a gapped phase to a gapless one, which is described as a spinless, interacting, 1D fermion system (Tomonaga-Luttinger liquid) [\onlinecite{Giamarchi_book}]. The filling of the spinless fermion band increases as a function of $H$ up to the second critical field $H_{\textrm{s}}$ corresponding to the complete filling of the band. The point ($H = H_{\textrm{s}}$, $T=0$) is the second quantum critical point separating the gapless TLL from a gapped ($\Delta$ = $g\mu_{\textrm{B}}(H-H_\textrm{s})$) fully polarized phase. This  scheme also applies to strong-leg coupling spin ladders, and  $S = 1$ spin chains (Haldane systems [\onlinecite{Haldane_conjecture}]), although their ground states and the gap are of different nature.
The same energy scheme applies to all quasi-2D or 3D systems of weakly coupled spin dimers,  but between $H_{\textrm{c}}$ and $H_{\textrm{s}}$ these systems are described as itinerant hard-core bosons on a lattice [\onlinecite{Giamarchi_natphys}], which ultimately condense at low temperature.  (c): Phase diagram of a spin ladder  in the $H-T$ plane. Above each of the two quantum critical points, there is a quantum critical regime in which the only energy scale is the temperature [\onlinecite{Mukhopadhyay_2012}]. For $E_{\rm{F}} \gg k_{\textrm{B}}T \gg J_{3D}$, where $E_{\rm{F}}$ is the Fermi energy of the spinless fermion band and $J_{3D}$ is the sum of the inter-ladder interactions, the system can be described as a TLL, while for  $J_{3D} > k_{\textrm{B}}T$, a 3D LRO is established, which can be described as a Bose-Einstein condensation of magnons [\onlinecite{GiamTsvelik1999}].  (d): Structure of a strong rung coupling spin ladder  (C$_{5}$H$_{12}$N)$_{2}$CuBr$_{4}$ (BPCB), from [\onlinecite{Klanjsek_2008}].}
\end{minipage}
\end{minipage}
\end{figure}

 Spin ladders are 1D systems made of two (or more) coupled spin chains. First discussed as a by-product of
 high $T_{\rm{c}}$ cuprates [\onlinecite{Hiroi,Dagotto}], they present a rich phase diagram in the $H-T$ plane [\onlinecite{Chaboussant_1998,Mila_98,Rice_Cargese}], as shown on Fig. \ref{Fig_spin_ladder_phase_diagram_v2} . We shall consider here only the most simple type of $S~=~1/2$ two-legs ladders, described by the following Hamiltonian:
 $\mathcal{H} = \sum_{i}^{l=1,2} J_{\|} \overrightarrow{S}_{i}^{l}.\overrightarrow{S}_{i+1}^{l}+J_{\bot}\sum_{i}\overrightarrow{S}_{i}^{1}.\overrightarrow{S}_{i}^{2}$
 where $J_{\|}$ ($J_{\bot}$) is the isotropic AF interaction along the legs (rungs). Whatever is the ratio $J_{\bot}/J_{\|}$, the ground state of two-legs spin ladders is the collective singlet separated by a gap from the first triplet excited states. As for the spin chains, the spin ladders  can be described in terms of spinless fermions. But here the filling of the band starts from zero at the magnetic field \Hc which closes the gap, up to the complete filling at \Hs. This is different from the case of the spin chains, where the filling starts always at half-filling at zero field. At temperature  high enough to neglect the inter-ladder couplings, and low enough as compared to the Fermi energy of the spin-less fermions, the low-energy excitations can be described in the framework of a TLL.
 Finding systems which are true spin ladders, with values of the AF couplings comparable to the  energy scale of the Zeeman coupling for fields accessible in the laboratory, is not so easy. An early candidate has been Cu$_{2}$(C$_{5}$H$_{12}$N$_{2}$)$_{2}$Cl$_{4}$ (called Cu(Hp)Cl) [\onlinecite{Chaboussant_1998}], presenting a phase diagram in the $H$-$T$ plane quite similar to that expected for a spin ladder , but the one-dimensionality of the system was questioned by neutron experiments [\onlinecite{Stone_2000}].
Recently, two spin ladder  compounds, well characterised by NMR, neutron spectroscopy and thermodynamic measurements, have been the subject of intense experimental and theoretical studies. The first one, (C$_{5}$H$_{12}$N)$_{2}$CuBr$_{4}$, usually called BPCB but also known as (Hpip)$_{2}$CuBr$_{4}$) [\onlinecite{Paytal,Watson}], is a strong rung coupling spin ladder  ($J_{\bot} \gg  J_{\|}$) with a \Hc~=~6.7~T and \Hs~=~11.9~T [\onlinecite{Klanjsek_2008}]. The whole phase diagram in the \textit{H}-\textit{T} plane was studied by NMR [\onlinecite{Klanjsek_2008}], specific heat [\onlinecite{Ruegg}], and neutron spectroscopy [\onlinecite{Thielemann}], and the results compared to Density Matrix Renormalisation Group DMRG and bosonization calculation using the values of $J_{\bot}$ and $J_{\|}$ derived from experiments.  The variation of the nuclear spin-lattice relaxation rate $1/T_1$ at fixed $T$ could be remarkably fitted in the whole interval \Hc to \Hs~ using a single parameter, the hyperfine coupling. Similarly, the field dependence of the transition temperature \Tc~of the 3D phase as well as that of its order parameter could be fitted using as a single parameter, the interchain coupling $J'$. Using time-dependent DMRG, even the high energy excitations  observed by inelastic neutron scattering could be very well reproduced, starting from the known values  $J_{\bot}$ and $J_{\|}$ of the Hamiltonian [\onlinecite{Bouillot}].

 The second spin ladder  compound, (C$_{7}$H$_{10}$N)$_{2}$CuBr$_{2}$ (called DIMPY) [\onlinecite{Shapiro}],  is in the strong-leg coupling regime ( $J_{\|} > J_{\bot}$) [\onlinecite{Hong,Schmidiger}] with \Hc~=~3~T and \Hs~=~29~T [\onlinecite{Jeong2016}]. This spin ladder  belongs to the regime where the quasi particles are attractive \mbox{($K \geq 1$)}. Although the determination of the TLL parameter $K$ from $1/T_1$ is difficult [\onlinecite{Jeong}], the variation of $1/T_1$ as a function of $H$ between \Hc~=~3~T and \Hs~=~29~T measured at constant temperature \mbox{$T$~=~750~mK} [\onlinecite{Jeong2016}] was perfectly described by the Luttinger liquid parameters $K$($H$) and $u$($H$) determined from the starting Hamiltonian where $J_{\|}$ and $J_{\bot}$ have been determined from neutron spectroscopy.

 In conclusion, spin ladders offer a rare example in which the TLL parameters can be computed directly from the microscopic parameters of the Hamiltonian. In that sense, they are perfect simulators of interacting spin-less fermions.

\subsubsection{The nematic phase in frustrated $J_1-J_2$ chain. The LiCuVO$_4$ compound.}
In spin systems, the frustration of the exchange couplings is known to lead to exotic ground states [\onlinecite{Frustration_book_2011}]. Here we describe the case of the spin nematic phase, for which the compound LiCuVO$_4$ seems to be the most promising system. LiCuVO$_4$ belongs to the class of the frustrated $J_1-J_2$ spin chains, in which the first neighbour exchange interaction $J_1$ is ferromagnetic (FM), while the next nearest neighbour is AF [\onlinecite{Enderle}]. In such a system, the saturated FM state at \Hs~was shown to be unstable with respect to the formation of pairs of bound magnons [\onlinecite{Hikihara,Sudan,Zhitomirsky}]. In their domain of stability, these bound magnons give rise to an SDW phase, in which the transverse fluctuations are gapped, and, just before the saturation to a nematic phase, in which the order parameter does not transform as a vector, but  as a quadrupolar tensor  of the type $S^{i}_{+}S^{i+1}_{+} + cc $. At lower field, single-magnon excitations prevail, giving rise to a vector chiral phase. The planar vector chiral phase and the longitudinal SDW attributed to the bound magnons have indeed been observed [\onlinecite{Enderle,Buttgen_2012,Nawa}] and here we focus only on the nematic phase. The search for this phase in LiCuVO$_4$ was triggered by the observation of an anomaly in the magnetisation curve just below the saturation field (45~T for $H\|c$, 52~T~for $H\|b$) [\onlinecite{Svistov}]. At such a high field, the only  available \textit{microscopic} technique is the  NMR. The first experiment was done for $H\|c$ in a steady magnetic field [\onlinecite{Buttgen_2014}], using the hybrid magnet at Tallahassee. It was concluded that the anomalous phase observed by magnetisation was essentially due to impurities, most of the system being saturated above 41.4~T, and that the nematic phase, if present, could only exist between 40.5~T~$< H < $~41.4~T. However, very recent experiments conducted in pulsed magnetic field for $H\|c$ and $H\|b$ have shown that, in both orientations, there is a field range below the saturation field in which there is no transverse order and the observed field dependence of the hyperfine shift is identical to the change in the magnetic susceptibility [\onlinecite{Orlova_2017,Mila_2017}], in agreement with the theoretical predictions for a nematic phase. Further evidence could be given by measuring $1/T_1$ as a function of temperature from the partially polarised phase down into the nematic one (at fixed value of $H$) [\onlinecite{Smerald}], but this type of measurement at such high magnetic fields are very challenging.
\subsection{Magnetisation plateaus in the quasi-2D Shastry-Sutherland compound SrCu$_{2}$(BO$_{3}$)$_{2}$}

The Shastry-Sutherland Hamiltonian (SSH) [\onlinecite{Shastry}]  describes a 2D network of orthogonal dimers, which applies to
SrCu$_{2}$(BO$_{3}$)$_{2}$ [\onlinecite{Kageyama1999}], the prototype
compound for the study of magnetisation plateaus. In the SSH, one
considers a square lattice that is paved by  orthogonal dimers
with an AF exchange $J$ along the diagonals (next nearest
neighbours), and then introduce an frustrating AF coupling $J'$
between each nearest neighbours.  The product of
singlets on the dimers is always an eigenstate of the SSH, whatever is the
ratio $J'/J$, but it remains the ground state only for $J'/J < 0.67$. For larger values of $J'/J$, it enters a narrow plaquette
phase [\onlinecite{Koga2000,Corboz2013}] before turning to a N\'eel
ground state. In the SrCu$_{2}$(BO$_{3}$)$_{2}$ compound $J'/J = 0.63$, so the
ground state is the product of singlets on each dimer. However,
the hopping of a triplet from one dimer to its nearest neighbour
is strongly restricted,  favoring  the existence of magnetisation
plateaus. The first evidence for their  existence  in
SrCu$_{2}$(BO$_{3}$)$_{2}$ was obtained by magnetisation
measurements in pulsed magnetic field [\onlinecite{Onizuka2000}] with the
observation of three plateaus at 1/8, 1/4 and 1/3 of the
saturation magnetization. The first microscopic insight of the
spin pattern of the 1/8 plateau was obtained soon after by Cu NMR
in a resistive magnet at the LNCMI
 [\onlinecite{Kodama2002,Takigawa2004}], opening a long term collaboration between the Grenoble NMR group with that of M. Takigawa at the Institute of Solid State Physics at Tokyo.

\begin{figure}[t]
\vspace*{-1.1cm}
\begin{minipage}{0.98\hsize}
\centerline{\includegraphics[scale=0.5]{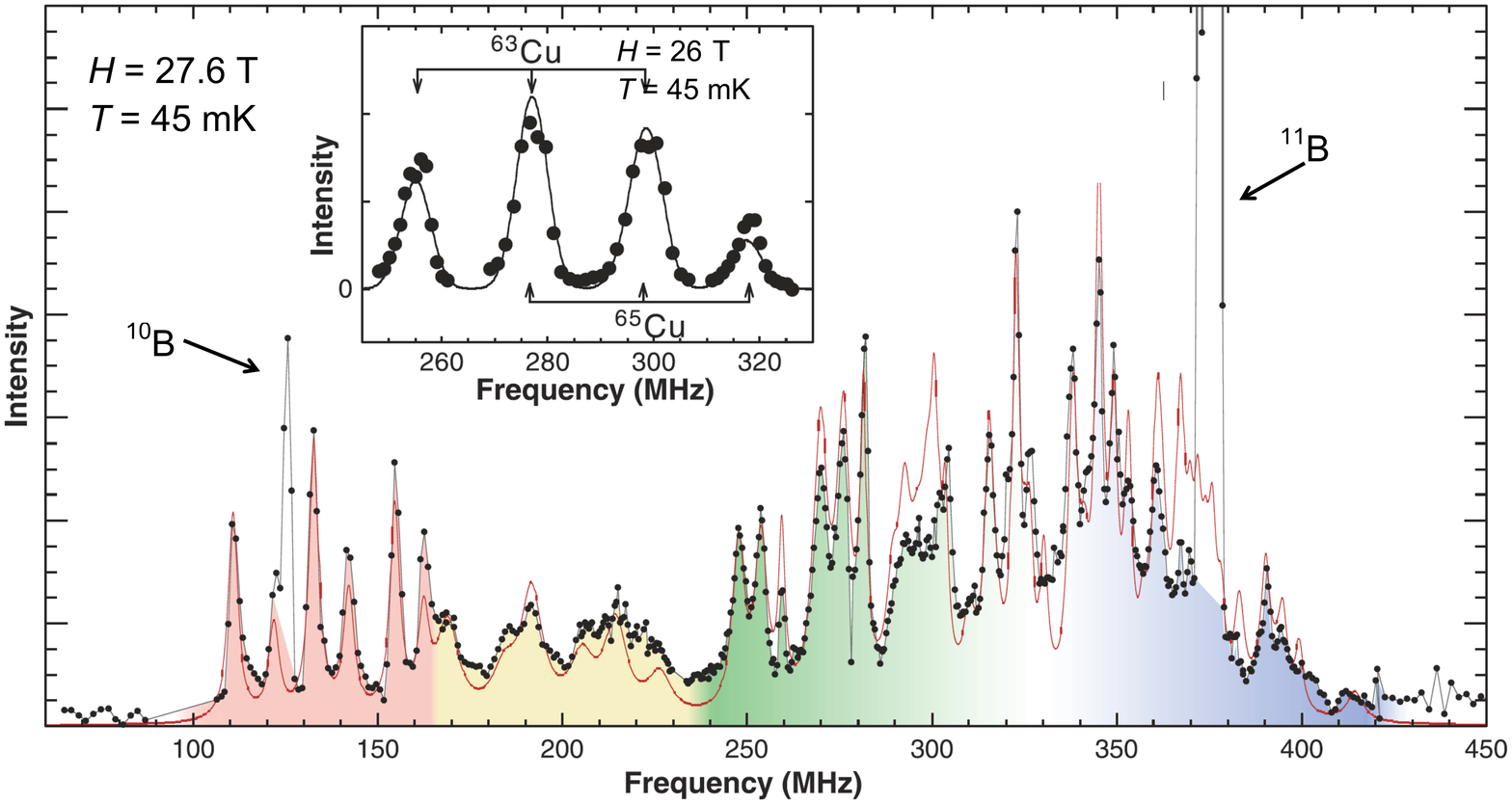}} 
\begin{minipage}{.98\hsize}
 \vspace*{-1.1cm}
\caption[]{\label{FigScience} (Color online)  Cu NMR spectrum in
the 1/8 plateau of SrCu$_{2}$(BO$_{3}$)$_{2}$. The spectrum extends
over $\simeq$ 300 MHz, corresponding to a range of $\simeq$ 30~T for the histogram of the hyperfine
fields. To each site correspond 6 lines due to the
two isotopes $^{63}$Cu and $^{65}$Cu and the quadrupole coupling,
which does not depend on the site. Note that the width
of stability of the plateau is only $\simeq$ 1.5~T which
precludes any field sweep at constant frequency to obtain the full
spectrum. From Ref. [\onlinecite{Kodama2002}].}
\end{minipage}
\end{minipage}
\end{figure}

As observed in the inset of Fig.~\ref{FigScience}, a single site of Cu (6 lines) is
observed as long as
 the magnetisation grows from zero in the gapped state to the
 plateau, meaning a uniform polarisation of the Cu$^{2+}$ electronic spins.
The main figure shows that inside the plateau  at least 10 different sites (60 lines)
 appear, spread over a distribution of internal field $^{63,65}A_{hyp}^{onsite}<S_z>$  of the order of 30~T. In particular, two strongly polarized sites are well resolved on the left of the spectra (the on-site hyperfine coupling for the Cu nuclei being strongly negative, Cu lines corresponding to strongly polarised Cu sites are strongly shifted to low frequency). Although the NMR spectra clearly demonstrate that the triplets crystalise within a commensurate super cell, they only give the histogram of the internal field due to these crystalised triplets but not the real magnetic structure inside the unit super cell, nor its symmetry. Exact diagonalisation of the  Heisenberg SSH on a 16 spins cluster led to the conclusion that the unit cell (16 Cu$^{2+}$ spins per layer) was rhombohedral, with oscillations of the spin polarization inside. Further calculations for the 1/4 and 1/3 plateaus showed the existence in all cases of strongly polarised ``dressed triplet" extending on three dimers, forming stripes.
\begin{figure}[h]
\begin{minipage}{0.98\hsize}
\centerline{\includegraphics[scale=0.9]{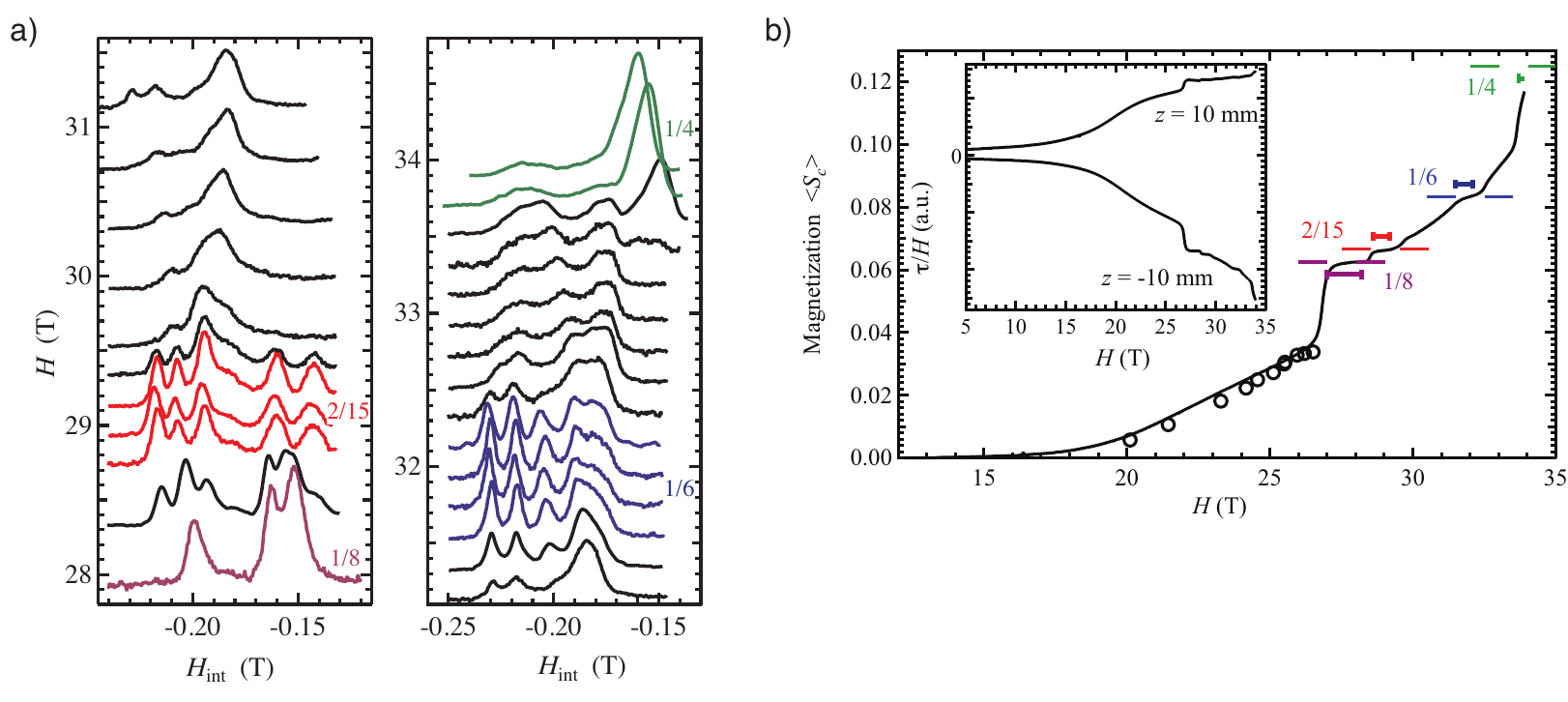}}
\begin{minipage}{.98\hsize}
\vspace*{+0.1cm}\caption[]{\label{Fig4plateaus}(Color online) Observation of the four plateaus at 1/8, %
2/15, 1/6 and 1/4 in \mbox{SrCu$_{2}$(BO$_{3}$)$_{2}$}. (a)
partial \mbox{$^{11}$B} NMR spectra between 28~T and 34~T. Within
the plateaus, the internal field histogram doesn't change with the
magnetic field, and the lines are sharp. (b) magnetization
measurement (main figure) obtained from torque measurements (inset). One clearly
observes the plateaus at  1/8, 2/15 and 1/6 of the magnetisation
at saturation $m_{sat}$.  From Ref. [\onlinecite{Takigawa2013}].}
\end{minipage}
\end{minipage}
\end{figure}

Further torque measurements [\onlinecite{Levy2008,Sebastian2008}] showed that
the plateau sequence was not limited to 1/8, 1/4 and 1/3.
While Ref. [\onlinecite{Levy2008}] reported the existence of a second
plateau adjacent to the 1/8  also observed through $^{11}$B
NMR, Ref. [\onlinecite{Sebastian2008}] claimed the existence of a full
series of plateaus at values of $\frac{m_z}{m_{sat}} = 1/q$
(2~$\leq~q~\leq 9$ )  and 2/9. This was followed by new theoretical
attempts to calculate the sequence of plateaus and their stability
[\onlinecite{Dorier2008,Abendschein2008}], rendered difficult by the
proximity of the quantum critical point  at $J'/J = 0.67$, by the
necessity to take into account long range interaction between the
triplets as well as the  additional terms to the SSH Hamiltonian like
Dzyaloshiskii-Moriya interaction [\onlinecite{Nemec2012}]. With
improvement of the pulse magnetic field set-ups, the 1/2 plateau
 was observed starting at 84~T [\onlinecite{Jaime2012}] and ending at 118~T
[\onlinecite{Matsuda2013}]. The exact sequence of plateaus up to the 1/4 one,
that is $\frac{m_z}{m_{sat}} = $ 1/8, 2/15, 1/6 and 1/4 was finally
established by $^{11}$B NMR up to 34~T [\onlinecite{Takigawa2013}],
combined with careful torque measurements as shown in Fig.~\ref{Fig4plateaus}b. One observes that within the field range corresponding to a magnetization plateau the internal fields stay constant (Fig.~\ref{Fig4plateaus}a), as expected in a gapped phase.  After deconvoluting the full
$^{11}$B spectra to get rid of the quadrupolar splitting
[\onlinecite{Mila2013}] and keep only the central lines shifted by the
internal field, structures were proposed for the 1/4, 1/6, 2/15
and 1/8 plateaus [\onlinecite{Takigawa2013}]. However,
a new theoretical approach (infinite projected entangled-pair
states, iPEPS)  finally established that the more stable
structures of the plateaus consist of triplet bound states with
$S_z = 2$ [\onlinecite{Corboz2013,Corboz2014}]. Employing NMR spectra to distinguish
between those new structure and the previous ones is delicate, because
of the effect of the dipolar field of the electronic
spins in the adjacent planes on the $^{11}$B spectra. A more direct approach
would be to repeat the Cu NMR spectra with a better control of the
intensity of the lines. Another possibility is a direct
neutron measurement at 26 T, provided the application of pressure
would lower enough the threshold field of the 1/8 plateau
[\onlinecite{Schneider2016,Zayed2017}]. %

\subsection{Bose Einstein Condensation of triplet excitations in coupled dimer systems.}
An important class of quantum AF systems  can be viewed  as a collection of dimers -~pairs of spin 1/2 strongly coupled by an AF exchange coupling $J$~-~on a quasi-1D, quasi-2D, or 3D lattice,  coupled by weaker interdimer interactions $J'$. These systems  have a collective singlet ground state separated by a gap $\Delta$ from triplet excited states, this gap being determined by a combination of $J$ and the interdimer  $J'$. These excitations, often called triplons, can be treated as hard-core bosons on a lattice. Their density at $T = 0$ is zero below the critical field \Hc, and above $H_{\textrm{c}}$ is controlled by the applied magnetic field which plays the role of a chemical potential. The hopping between neighbouring sites is controlled by $J'$.
Although such a description was used long time ago to describe the superfluid properties of Helium \onlinecite{Matsubara1956}, it is only in 1999-2000 that the quest for Bose-Einstein condensation started in quantum antiferromagnets \onlinecite{GiamTsvelik1999,Nikuni2000}, opening a wide area of research \onlinecite{Giamarchi_natphys,Zapf2014}. We shall not enter in details in that field (references can be found in [\onlinecite{Giamarchi_natphys,Zapf2014}]) but concentrate on the NMR point of view. The condition to obtain a BEC in a QSS is the invariance of the spin Hamiltonian under a rotation around the applied magnetic field. At the onset of the BEC, a transverse ($\bot H_0$) staggered magnetization appears, which is the order parameter of the BEC. The transition can be observed as a function of the temperature, or as a function of the magnetic field at the quantum critical point $H_{\textrm{c}}$. For $H$ slightly larger than $H_{\textrm{c}}$ (or slightly smaller than the saturation field $H_{\textrm{s}}$), the hard-core bosons are very dilute, and one expects the transition temperature $T_{\textrm{BEC}}$ to vary as $(H-H_{\textrm{c}})^{\alpha}$. The exponent $\alpha$ is equal to $2/d$ where $d$ is the dimensionality of the system (usually $d = 3$).

From the NMR point of view, this transverse staggered magnetisation will split each NMR line of the paramagnetic phase into two lines, the separation of which is proportional to the order parameter. Since NMR measures only the projection of the hyperfine field along $H_0$, the observation of this splitting requires that the hyperfine tensor components $A_{zx}, A_{zy}$ are different from zero.  Such a splitting was first observed in TlCuCl$_3$ [\onlinecite{Vyaselev2004}], but the transition at $H_{\textrm{c1}}$ was found to be a first order one  accompanied by a lattice distortion. Better examples can be found in  spin ladders compounds [\onlinecite{Klanjsek_2008,Jeong}] and in the S = 1 spin  chain NiCl$_2$-4SC(NH$_2$)$_2$ (DTN) [\onlinecite{Zapf2006,Yin2008,Blinder2017}] in which $\alpha$ = 2/3 close to \Hc and \Hs.

We shall now consider the case of the compound BaCuSi$_2$O$_6$ [\onlinecite{Jaime2004}], which has drawn a lot of interest for its peculiar low temperature properties. In this quasi-2D compounds, the dimers are positioned perpendicular to the $a$-$b$ plane, forming a body-centered tetragonal lattice. It was reported  that below 880~mK and close to $H=H_{c}$, $T_{\textrm{BEC}}$ was not varying as $(H-H_{c})^{2/3}$, but as $H-H_{c}$  [\onlinecite{Sebastian2006}]. This linear dependence corresponds to a 2D BEC, and the 2D character was explained by invoking the frustration between adjacent planes due the body-centered structure [\onlinecite{Sebastian2006}]. Soon after, it was shown that the system was undergoing a structural phase transition at 90~K [\onlinecite{Samulon}], giving rise to two alternating types of planes (A and B) with different gaps in the magnetic excitations, implying two different critical fields $H_{\textrm{cA}} < H_{\textrm{cB}}$, and to an incommensurate distortion within the planes [\onlinecite{Ruegg2007,Kraemer2007}]. NMR experiments, performed between 13 and 26~T and at temperature as low as 50~mK,  allowed to determine the ratio of the gaps in the two types of planes $\Delta_\textrm{B}/\Delta_\textrm{A}$~=~1.16, provided an accurate determination of $H_{\textrm{c}}$ = 23.4~T, and confirmed the linear dependence of $T_{\textrm{BEC}}$ with $H-H_{\textrm{c}}$ in the low temperature range. They also showed that the triplon populations of the A and B planes were very different. An accurate determination of the variation the boson population $n_\textrm{B}$ in the plane B  with $H-H_{\textrm{cA}}$ was crucial to discriminate between different theoretical models and to explain the linear dependence of $T_{\textrm{BEC}}$ [\onlinecite{Rosch,Laflorencie2009}].  Further NMR experiments have been done on this purpose in a $^{29}$Si-enriched  sample. Since  the average longitudinal magnetisation, and hence the first moment of the NMR lines, is proportional to the boson population, it was possible to measure accurately the total boson population $n_\textrm{A} + n_\textrm{B}$ as a function of $H-H_{\textrm{cA}}$  from the first moment of the $^{29}$Si line, and the B planes boson population $n_\textrm{B}$  from the first moment of the $^{63}$Cu line, which is a very sensitive probe [\onlinecite{Kraemer2013}]. It was concluded that none of the models considering a perfect frustration could explain the very weak population observed in the B planes, and that the frustration between adjacent planes should be slightly released.
The story could have stopped here, but LDA+U calculations of the exchange couplings [\onlinecite{Mazurenko}] eventually showed that the effective coupling between adjacent dimers were ferromagnetic (in agreement with neutron data [\onlinecite{Ruegg2007}]), thus completely suppressing the frustration and radically changing the nature of this system. A new comprehensive theoretical description of this mysterious compound, including the (linear) dependence of $T_{\textrm{BEC}}$ with $H-H_{\textrm{c}}$, that of $n_\textrm{A}$ and $n_\textrm{B}$,  and the complete phase diagram in the $H$-$T$ plane, remains to be done.
\section{\label{HTSCsec}Field-Induced Charge Density Waves in cuprates High $T_{\rm{c}}$ superconductors.}

Thirty years after its discovery by Bednorz and M\"uller [\onlinecite{Bednorz}], the mystery of high temperature superconductivity in the cuprates has still not been cracked~[\onlinecite{HTCReview}]. Let us only recall here that the essential ingredient in the structure of these compounds is the CuO$_2$ plane, in which superconductivity takes place. These planes alternate with some charge
reservoirs from which doped holes are transferred. The typical phase diagram of these compounds starts from an AF state at zero hole
doping. Superconductivity appears above a hole doping level $p\simeq0.05$ but (glassy-type) magnetic order persists at low temperature over some material-dependent doping range (typically up to $p=0.08$ in YBa$_2$Cu$_3$O$_{y}$~[\onlinecite{Wu13}] and refs. therein). The superconducting temperature $T_c$ forms a dome with a maximum at $p\simeq0.16$. Its low (high) doping side is called the underdoped (overdoped) regime. Normal state electronic properties in the underdoped regime are strongly influenced by the presence of the celebrated pseudogap~[\onlinecite{HTCReview,Alloul89}].

Ten years ago, high field experiments have discovered quantum oscillations in the underdoped regime of YBa$_2$Cu$_3$O$_{y}$, and it was found that their frequency was much lower than in the overdoped regime [\onlinecite{HTCReview,QO,Vignolle2013}], indicating a much smaller Fermi surface volume in the underdoped regime. This, combined with the change of sign of the Hall effect~[\onlinecite{LeBoeuf2007}], hinted at a reconstruction of the Fermi surface in underdoped YBa$_2$Cu$_3$O$_{6+y}$, with electron pockets occupying only a few \% of the Brillouin zone [\onlinecite{Vignolle2013,Sebastian_2015}]. In order to determine the origin of this reconstruction, NMR experiments were undertaken in conditions of temperatures and magnetic fields and on single crystals of YBa$_2$Cu$_3$O$_{y}$ ($y \simeq 6.5$) comparable to those in which quantum oscillations had been observed [\onlinecite{Tao2011}].

These NMR experiments unambiguously demonstrated the presence of a CDW, without any concomitant spin order~[\onlinecite{Tao2011}].  Although NMR is a local probe, the observation of a line splitting, instead of a simple broadening, immediately suggested that the CDW order is coherent over fairly large distances, thus establishing the second case of long-range charge order in the cuprates, after the stripe phase observed in the La$_{2-x}$Sr$_{x}$CuO$_4$ family~[\onlinecite{Tranquada,Comin2016}]. Static CDW had also been identified by scanning-tunneling-microscopy (STM) in Bi-based cuprates but the correlations were quite short-ranged: $\xi_{\textrm{CDW}} \leq 2\lambda$ where CDW the period $\lambda$ is 3 to 4 lattice spacings~[\onlinecite{Comin2016}].

Oxygen-ordered YBa$_2$Cu$_3$O$_{y}$ being by far the ``cleanest" (the least disordered)  cuprate superconductor, the observation of long-range CDW ordering led to conclude that CDW has to be a generic tendency of underdoped cuprates, although long-range ordering may eventually not be achieved in most cases~[\onlinecite{Tao2011}]. This initial experiment also clearly established that CDW order competes with superconductivity since the effect is present only when $T_c$ is significantly reduced by fields applied perpendicular to the CuO$_2$ planes, and not for low fields or for high fields applied parallel to the planes, two situations in which superconductivity remains strong~[\onlinecite{Tao2011}]. Therefore, the appearance of long-range CDW order should not be viewed as an exotic field-induced phenomenon but instead the direct consequence of the suppression of superconductivity by high fields applied perpendicular to the CuO$_2$ planes. Indeed, it was recently demonstrated that the upper critical field $H_{c2}$ is severely reduced at doping levels where CDW order is observed~[\onlinecite{Grissonnanche,Zhou17b}].

The observation of a line splitting in the original $^{63}$Cu NMR data strongly suggested that the CDW is uniaxial in high fields~[\onlinecite{Tao2011}]. Two possible interpretations were mentioned: a modulation along the chain direction ($b$ axis) or a modulation perpendicular to the chains ($a$ axis). Because of similarities with the stripe phase around 1/8 doping in La-based cuprates, and because a modulation along the $a$ axis was also  independently detected in the YBa$_{2}$Cu$_{3}$O$_{6.54}$ sample (Ortho-II) (the splitting was observed for nuclei below full chains but not for those below empty chains), a $4a$-period modulation was favored. Later experiments using $^{17}$O NMR fully characterized the field dependence of the CDW due to its competition with superconductivity and established the presence of an onset field proportional to the upper critical field~[\onlinecite{Tao2013}]. An interpretation of the onset field in terms of critical density of halos of CDW order centered around vortex cores was proposed [\onlinecite{Tao2013}].

This founding NMR paper [\onlinecite{Tao2011}] was followed by an avalanche of X-ray studies [\onlinecite{Ghiringhelli2012,Chang2012,Blanco2014,Comin2014,SilvaNeto2014,Tabis2014,Comin2016}]. These quickly established  that CDW modulations are  indeed ubiquitous in underdoped cuprates and competing with superconductivity. However, in contrast with the high-field NMR results, the modulations were found to have a much higher onset  temperature, and to have two wave vectors ($Q,0$) and ($0,Q$), without any magnetic field threshold.
Furthermore, the coherence length of the modulation, $\xi_{\textrm{CDW}}$, was found to be relatively short $\xi_{\textrm{CDW}} \leq 5\lambda$ in \YBCO~ [\onlinecite{Blanco2014}], with the CDW period $\lambda$ equal to 3 to 4 lattice spacings, and very short, $\xi_{\textrm{CDW}} \leq 2\lambda$, in \Bi2212 [\onlinecite{Comin2014,SilvaNeto2014}] and \Hg1201 [\onlinecite{Tabis2014}].

Actually, this "normal state" short-range CDW is also detected by NMR~[\onlinecite{Tao2015}]. Qualitatively speaking, the situation is reminiscent of the short-range CDW order in the form of generalized Friedel oscillations,  observed by NMR [\onlinecite{Berthier78,Ghoshray2009}] and STM [\onlinecite{Arguello2014,Chatterjee}] around impurities  in NbSe$_2$. In NbSe$_2$, these oscillations are related to the (real part of) the dynamic susceptibility $\chi_{\rm{CDW}}$ of the pure system, in the same way as the screening of non-magnetic impurities in cuprates is related to the antiferromagnetic susceptibility~[\onlinecite{Julien2000}] or in the same way as the short-range, static nematic order in Fe-based superconductors is a consequence of disorder and of the existence of a large nematic susceptibility~[\onlinecite{Iye}]. In YBa$_2$Cu$_3$O$_{y}$, however, disorder arises mostly from out-of-plane oxygen defects, which presumably makes a weak pinning picture with phase defects and CDW domains more appropriate than the strong pinning picture with single point-like defects (see a related discussion from the x-ray perspective in ref.~[\onlinecite{LeTacon}]).

\begin{figure}[t]
\begin{minipage}{0.98\hsize}
\centerline{\includegraphics[scale=0.5]{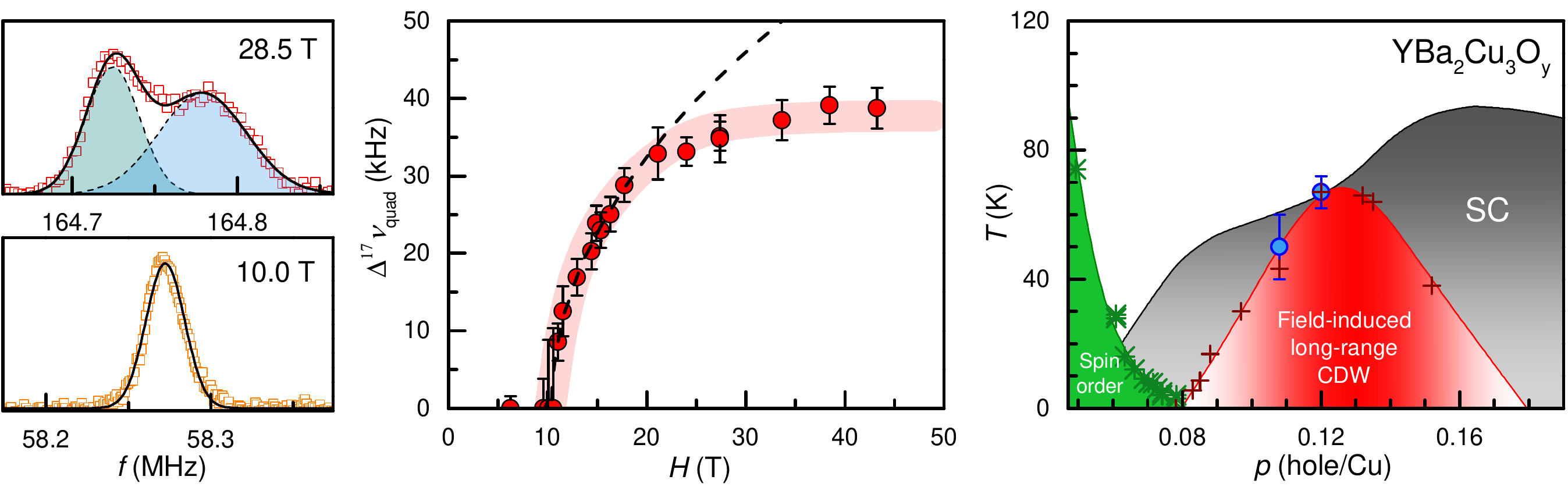}}
\begin{minipage}{0.98\hsize}
\caption[]{\label{FigCDWHTC}(Color online) NMR evidence for Field
Induced Charge density Wave in \YBCO. Left panels: Splitting of the $^{17}$O NMR line at $T\simeq 2$~K~[\onlinecite{Tao2013}]. Middle panel: Field dependence of the quadrupole part of the $^{17}$O line splitting~[\onlinecite{Tao2013}]. The dashed line represents a $\sqrt{H-H_{charge}}$ behavior close to the onset field $H_{charge}$. Right panel: Revised phase diagram of YBa$_2$Cu$_3$O$_{y}$, including the high field CDW from ref.~[\onlinecite{Tao2011}]. Crosses correspond to the sign change of the Hall effect and the blue points mark the temperature onset of the NMR splitting.}

\end{minipage}
\end{minipage}
\end{figure}

The relationship between the field-induced CDW, observed by NMR [\onlinecite{Tao2011,Tao2013}] and sound velocity measurements [\onlinecite{LeBoeuf2013}],  and the zero-field short-range CDW, observed by X-rays [\onlinecite{Comin2016}] and NMR [\onlinecite{Tao2015}],  has puzzled the community for a while. However, the presence of two distinct, albeit obviously related, CDW phases was eventually accepted when field-induced long-range CDW was confirmed by X-ray experiments in high magnetic fields [\onlinecite{Xraypulsedfields,Chang2016,Jang2016}]. These experiments showed that the CDW is indeed of single-$Q$ type, but along the $b$ axis and incommensurate. They also showed that the NMR threshold field corresponds to a growth of the correlation length $\xi_{\textrm{CDW}}$ in the CuO$_2$ planes. Basically, $\xi_{\textrm{CDW}}$  becomes large enough to produce an NMR pattern --the histogram of the frequency distribution due to the charge modulation-- typical of a single $Q$ CDW [\onlinecite{Blinc81,Butaud1995}]. On the other hand, the thermodynamic transition observed by the ultrasound technique~[\onlinecite{LeBoeuf2013}] occurs at slightly higher field (and, presumably, a slightly lower temperature for the same field), marking the onset of coherence along the $c$ axis. The observation of a 2D pattern in NMR does not require phase coherence along the $c$-axis, as long as the phase fluctuations in the planes are pinned. The long-range CDW phase stacks in-phase along $c$, while the short-range CDW stacks out-of-phase. Both NMR and X-rays find that short-range correlations remain within the long-range phase~[\onlinecite{Tao2015,Xraypulsedfields,Chang2016,Jang2016}]. This is likely related to the presence of disorder [\onlinecite{Tao2015,Jang2016,Julien15}].

Addressing all the recent developments related to CDW order in the cuprates is beyond the scope of this short review on high-field NMR (see refs.~[\onlinecite{HTCReview,Comin2016,Julien15,PDW}] for recent perspectives). Let us only mention here the discovery of an intra-unit-cell d-wave symmetry of the CDW as one of the outstanding recent achievements~[\onlinecite{Fujita14,Comin15a}], showing that this CDW is by no means conventional. Many outstanding questions are still debated, such as whether the Fermi surface is primarily reconstructed by the short-range or the long-range CDW, the role of disorder in shaping the complex phenomenology, and, most importantly, the relationship between CDW and other phenomena in the pseudogap phase (particularly other ordering phenomena) and  the relationship with superconductivity. NMR investigation of the field-induced CDW in YBa$_2$Cu$_3$O$_{y}$ is being pursued~[\onlinecite{Zhou17}], other systems will be investigated with high-field NMR (see a puzzling recent report in Bi$_{2}$Sr$_{2-x}$La$_x$CuO$_{6+\delta}$~[\onlinecite{Kawasaki17}]) and it is both desirable and likely that other techniques like Raman scattering and optical spectroscopy will provide new insights upon going to high fields. Clearly, this research area will benefit from further development of experiments in, always higher, pulsed and steady magnetic fields.

\section{\label{exoticSC}Exotic superconductivity}

\subsection{FFLO state in $\kappa$-(BEDT-TTF)$_2$Cu(SCN)$_2$}
The Fulde-Ferrell-Larkin-Ovchinikov (FFLO) [\onlinecite{Fulde,Larkin}]
state is expected to occur in the vicinity of the upper critical
field ($H_{\textrm{c2}}$) when Pauli pair breaking dominates over orbital
effects [\onlinecite{Maki,Gruenberg}]. Pauli pair breaking prevails in
fields  for which the Zeeman energy is strong enough to break the
Cooper pair by flipping one spin of  the singlet. In a FFLO
state, the Copper pairs acquire a finite momentum, leading to a
modulated superconducting state, which can be schematised as periodically alternating ``superconducting" and ``normal" regions.
In solid state physics, the search for FFLO state has been mainly focused
on the heavy fermion compound CeCoIn$_5$ [\onlinecite{Bianchi03,Koutroulakis10}] and layered organic
superconductors [\onlinecite{Singleton00,Lortz07,Bergk11,Uji06}]. In the
case of CeCoIn$_5$, the phase initially identified as an FFLO one has been shown to be magnetically ordered
[\onlinecite{Young2007,Kenzelman2008}], and the putative coexistence with an FFLO state is still a matter of debate
[\onlinecite{Koutroulakis10,Hatakeyama2015,Kim2016}].

\begin{figure}[h]%
\begin{minipage}{0.98\hsize}
\centerline{\includegraphics[scale=0.6]{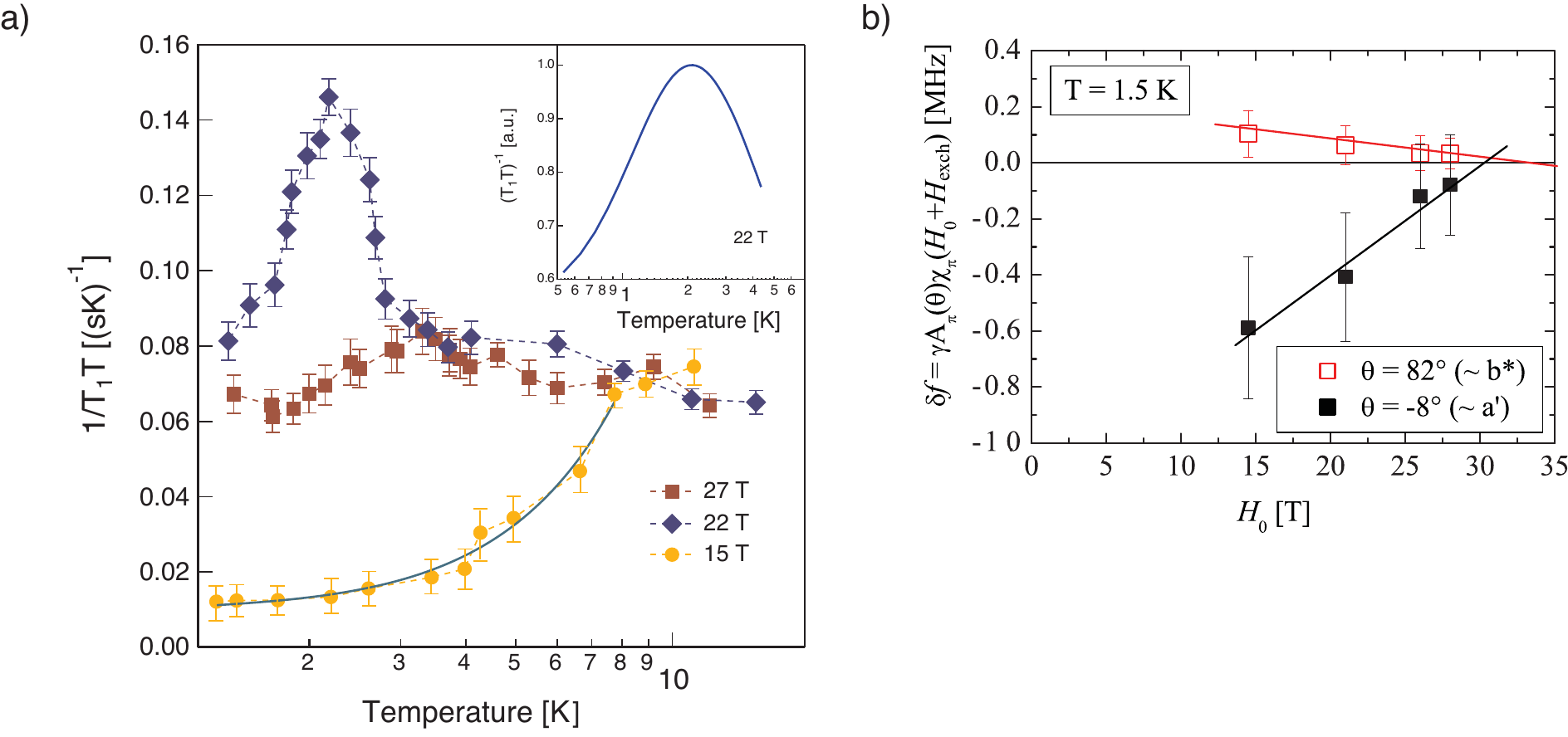}}

\begin{minipage}{0.98\hsize}
\caption{\label{FFLO}
(Color online) (a): NMR relaxation rate in the normal and superconducting states of $\kappa$-(BEDT-TTF)$_2$Cu(SCN)$_2$.
Temperature dependence of $^{13}$C NMR $(T_1T)^{-1}$ at fields of 15, 22 and 27 T,
applied in the conducting planes (symbols). In agreement with the phase diagram based on magnetic torque measurements [\onlinecite{Bergk11}], at 15~T, the system becomes superconductor around 7~K (Solid line denotes the quadratic temperature dependence characteristic for superconductors with a gap having a line of nodes, such as for a $d$-wave symmetry).  At 27~T, the system remains normal down to the lowest temperature investigated (1.4~K), while at 22 T, the system enters from the normal state into the FFLO phase. The huge peak in $(T_1T)^{-1}$ is due to the Andreev bound states present in the FFLO phase [\onlinecite{Mayaffre14,Vorontsov2005,Vorontsov2016}]. Inset: Simulation of the $(T_1T)^{-1}$ peak due to the Andreev bound states. (b): Field dependence of the spin polarization of the conduction electrons of the $p$-band as measured by $^{77}$Se NMR  (see the text) for two different orientations of the field in the superconducting plane of $\lambda$-(BETS)$_2$FeCl$_4$. The two straight lines cross at $H_0$~=32~$\pm$~2~T, when $H_0+H_{\textrm{exch}}$~=~0, in agreement with the maximum of \Tc~observed at 33~T [\onlinecite{Balicas_2001}].}
\end{minipage}
\end{minipage}
\end{figure}

 Quasi-2D organic superconductors are indeed ideal systems to observe FFLO state,
because of their large anisotropy: when the magnetic field is strictly aligned within the superconducting planes, there are only Josephson
vortices [\onlinecite{Roditchev}] left. Thermodynamic measurements have shown in the phase diagram of the compound $\kappa$-(BEDT-TTF)$_2$Cu(SCN)$_2$
 the existence of a narrow additional superconducting phase just below $H_{\textrm{c2}}$, for $H$ aligned with the
conducting plane [\onlinecite{Lortz07,Bergk11,Agosta2012,Agosta2016}], which could be an FFLO phase. The first high-field NMR experiment
[\onlinecite{Wright}] was performed at 0.35 K and concluded that the phase transition observed at 21.3~T was Zeeman driven.
In spite  of efforts to observe directly the spatial modulation of the order parameter, it has not been seen yet. However, it was noted  that, due to the modulation of the order parameter, nodes occur forming domain walls in which the superconducting phase changes by $\pi$ [\onlinecite{Vorontsov2005}]. This phase twist leads to a local modification of the density of states and the creation of new topological defects, characterized by the formation of Andreev bound states (ABS), which are a hallmark of the FFLO phase.
Recently, a high-field NMR experiment [\onlinecite{Mayaffre14}] has shown that these spin-polarized ABS produce a huge enhancement of the NMR relaxation rate $1/T_1$ (Fig~\ref{FFLO}a), providing the first microscopic characterisation of an FFLO phase. It turns out that this effect only occurs in a limited range of relatively high temperature, which has been consistently explained by theory [\onlinecite{Vorontsov2016}], showing that this enhancement comes from the scattering of electronic spins between the bound and continuum states. A new compound with  much smaller $H_{\textrm{c2}}$ values, $\mbox{$\beta$"-(BEDT-TTF)$_2$SF$_{5}$CH$_{2}$CF$_{2}$SO$_3$}$, has been recently investigated [\onlinecite{Koutroulakis16}] by NMR, confirming the nuclear spin-lattice relaxation results obtained in [\onlinecite{Mayaffre14}] are ubiquitous of the FFLO phase, and observing a lineshape compatible with a single-$q$ modulated superconducting phase.


\subsection{Field induced superconductivity in $\lambda$-(BETS)$_2$FeCl$_4$}
To explain field-induced superconductivity, Jaccarino and Peter [\onlinecite{Vincent}] have proposed a
mechanism  in which there is a compensation mechanism between the
applied magnetic field and the effective negative field seen by
the conduction electrons through an exchange mechanism with
polarised localised spins. The best realisation of this phenomenon
happens in the organic compound $\lambda$-(BETS)$_2$FeCl$_4$, which
is a charge transfer salt composed of the organic BETS
(C$_{10}$S$_{4}$Se$_{4}$H$_{8}$, bisethylenedithiotetraselenafulvalene) donor molecule
and magnetic FeCl$_4$ (Fe$^{3+}$, $S = 5/2$) counter ion
[\onlinecite{Kobayashi}]. At $H = 0$, the  localised Fe spins order below
8~K and the compound becomes insulator. Above 12 T, the AF order is suppressed, the system becomes metallic, and the Fe spins
$\mathbf{S}$ are fully polarized at sufficiently low temperature.
Increasing $H$ further, a superconducting phase appears at $H =
18-20$ T (depending of the orientation of the field) which has to
be strictly confined in the $b^{*}$-$c$ BETS conducting plane, thus suppressing the orbital limit
 [\onlinecite{Uji_2001}]. Because of the presence of the localized spins $\mathbf{S}$, the only technique allowing to measure the
 polarisation of conduction electrons is NMR. The experiment was performed on a  $^{77}$Se enriched, tiny single crystal of
 dimensions 3~x~0.05~x~0.01~mm$^3$, placed in an NMR coil of 70~$\mu$m diameter [\onlinecite{Hiraki_2007}].
 The  Hamiltonian  describing the interaction between the nuclear spins $\mathbf{I}^{i}$, the conduction electrons $\mathbf{s}^{k}$ of the conduction band,
 and the localized spins $\mathbf{S}^{j}$ can  be written as
 \begin{equation}
 \mathcal{H} = \mathcal{H}_Z + \mathcal{H}_{Is} + \mathcal{H}_{IS}
 + \mathcal{H}_{exch}.
 \end{equation}
 The first term is the Zeeman interaction
 \begin{equation}
\mathcal{H}_Z = \sum_i -\gamma\hbar I_z^i(1+K_c^i)H_0 + \sum_j
g_{Fe}\mu_{\rm{B}}S_z^j H_0 + \sum_k g_{\pi}\mu_{\rm{B}} s_z^k~,
 \end{equation}
in which $K_c^i$ is the chemical shift, $g_{\pi}$ the $g$ factor
for the $\pi$ electron of the conduction band. The second and the third terms are respectively the hyperfine interaction of the $p$-band and the dipolar interaction between the nuclear and the localised Fe spins. This latter can be calculated exactly, taking into account the
demagnetisation field.
 Concerning the last term, experiments were conducted at sufficiently low temperature and high magnetic field
so that the magnetisation of the Fe ions was saturated and field independent. $\mathcal{H}_{exch}$ can thus be simply written
$\mathcal{H}_{exch} \approx \sum_k g_{\pi}\mu_{\rm{B}} s_z^k H_{exch}$. Finally, the shift from the Larmor frequency, after
removal of the dipolar contribution and the chemical shift can be written as
\begin{equation}
\delta f^i =f^i - \gamma H_0 =
  \gamma[A_{\pi}^i(\theta)\chi_{\pi}(H_0+H_{exch})],
\end{equation}
which is a linear function of $H_0$. For two different orientations
$\theta$, the lines $\delta f$($H_0$) cross zero at the same field value (Fig.~\ref{FFLO}b),  allowing the
determination of the (negative) value of $H_{exch} = -32 \pm 2$~T [\onlinecite{Hiraki_2007}],
 in excellent agreement with the value of 33 T corresponding to the maximum of \Tc~ [\onlinecite{Balicas_2001}].

\section{Conclusion}
In this  review paper, we have shown the interest of performing NMR in very high magnetic fields, to explore new field-induced quantum ground states in condensed matter. We have limited ourselves to the case of quantum magnetism, high \Tc~superconductors and exotic superconductivity, but many other fields can be considered, like heavy fermions or Dirac electrons for example. There are, of course, some limitations of NMR with respect to other techniques: only  nuclei with a sufficient isotopic abundance and suitable gyromagnetic ratio can be studied (although the former constrain can be escaped by isotopic enrichment). On the other hand, tiny samples can be studied, which is not always the case for neutron inelastic scattering. The paper is focused on NMR experiments performed in very high field resistive magnets, but the physics as a function of the magnetic field must be considered as a whole, and the separation between the use of superconducting magnets, resistive magnets, and pulsed magnetic fields is purely technical. In any case, experiments using the two last field sources should be carefully prepared at lower field in superconducting magnets, which are less expensive, and for which the duration of experiments is not limited. We note that fields accessible with superconducting magnets devoted to solid state  physics have recently reached 24.6 T [\onlinecite{IMR}]. Obviously, the development of NMR in high magnetic field relies on pushing this limit as high as possible. Up to recently, NMR was the only technique allowing to get microscopic information above 17~T. Nowadays, the development of a dedicated hybrid magnet for neutron scattering provides steady magnetic field up to 27~T, and  X-rays scattering under pulsed magnetic field, will allow a fruitful comparison between all these complementary techniques.
\section*{Acknowledgements}

We are indebted to all the participants to the work reported here, in particular  Y. Fagot-Revurat, M. Klanj\v{s}ek,  M. Jeong, R. Blinder, M. Grbi\'c, A. Orlova, M. Takigawa and R. Stern for experiments in quantum spin-systems, T. Wu, R. Zhou, M. Hirata, I. Vinograd, A. Reyes, P. Kuhns for experiments in high \Tc~superconductors and  V.E. Mitrovi\'c, K. Miyagawa and K. Hiraki for those performed in exotic superconductors. We also acknowledge T. Giamarchi, F. Mila, N. Laflorencie for their theoretical support. Most of the works presented here were supported by the European Commission Contracts No. RITA-CT-2003-505474 (High Field research), RII3-CT-2004-506239 (EuroMagNet) and FP7-INFRASTRUCTURES-228043 (EuroMagNet II); by the French ANR Grants No. ANR-06-BLAN-0111 (NEMSICOM), ANR-12-BS04-0012 (SUPERFIELD) and ANR-14-CE32-0018 (BOLODISS); by the Laboratoire d'excellence LANEF in Grenoble (ANR project No. 10-LABX-0051); and by the p\^ole SMINGUE of the Universit\'e Grenoble Alpes.

\end{document}